\documentclass[onecolumn,superscriptaddress,prl,nofootinbib]{revtex4}
\usepackage{graphicx}
\usepackage{color}
\usepackage{epsfig,subfigure}
\usepackage{amsmath,amssymb,latexsym}
\usepackage{ascmac}
\usepackage{eclbkbox}

\def\U#1{{\rm #1}} 

\newcommand{\bra}[1]{\langle #1 |}
\newcommand{\ket}[1]{| #1 \rangle}

\newcommand{\expect}[1]{\left\langle #1 \right\rangle}

\begin{document}
\title{
Finite-key security analysis of quantum key distribution with imperfect light sources}
\author{Akihiro Mizutani$^*$}
\affiliation{Graduate School of Engineering Science, Osaka University,
Toyonaka, Osaka 560-8531, Japan}
\author{Marcos Curty}
\affiliation{EI Telecomunicaci\'on, Department of Signal Theory and Communications, University of Vigo, Vigo E-36310, Spain}
\author{Charles Ci Wen Lim}
\affiliation{Group of Applied Physics, University of Geneva, Geneva CH-1211, Switzerland}
\author{Nobuyuki Imoto}
\affiliation{Graduate School of Engineering Science, Osaka University,
Toyonaka, Osaka 560-8531, Japan}
\author{Kiyoshi Tamaki}
\affiliation{NTT Basic Research Laboratories, NTT Corporation, 
3-1, Morinosato-Wakamiya Atsugi-Shi, 243-0198, Japan\\
${}^*$mizutani@qi.mp.es.osaka-u.ac.jp}
\begin{abstract}
In recent years, the gap between theory and practice in quantum key distribution (QKD) has been significantly narrowed, 
particularly for QKD systems with arbitrarily flawed optical receivers. 
The status for QKD systems with imperfect light sources is however less satisfactory, 
in the sense that the resulting secure key rates are often overly-dependent on the quality of state preparation. 
This is especially the case when the channel loss is high. 
Very recently, to overcome this limitation, Tamaki {\it et al} proposed a QKD protocol based on the so-called 
``rejected data analysis'', and showed that its security---in the limit of infinitely long keys---is almost independent 
of any encoding flaw in the qubit space, being this protocol compatible with the decoy state method. 
Here, as a step towards practical QKD, we show that a similar conclusion is reached in the finite-key regime, 
even when the intensity of the light source is unstable. 
More concretely, we derive security bounds for a wide class of realistic light sources and show that the bounds 
are also efficient in the presence of high channel loss. 
Our results strongly suggest the feasibility of long distance provably-secure communication with imperfect light sources.
\end{abstract}

\maketitle
\section{I. introduction}
The gist of quantum key distribution (QKD) \cite{Gisin2002,scarani09,lo14} is that it allows two remote parties, 
Alice and Bob, 
to establish common secret keys in the presence of an adversary, Eve, who may have unlimited computing resources 
and technological advances. 
Today, three decades after its introduction, QKD has made enormous progress in both theory and practice, 
and is arguably on the verge of global commercialisation.~Having said that, however, there are still some issues, 
both theoretical and experimental, that need to be resolved before we can reach that level. 
Amongst those, the most pressing one is the mismatch between device models used in security proofs and actual devices used 
in QKD systems. 
In particular, such implementation loopholes can lead to side-channel attacks that break the security of QKD. 
Notably, it has been repeatedly demonstrated that the behaviour of single-photon detectors employed in QKD systems can be 
externally controlled, simply by exploiting their physics~\cite{hack:d}. 
In this case, it is easy to verify that security cannot be achieved, since the measured data are not representative of the 
quantum channel~\cite{Lim2015}. 
Undoubtedly, such hacking demonstrations raise not only the importance of proper calibration of QKD systems, 
but also the importance in developing security proof techniques that can tackle modeling discrepancies. 
Indeed, in the past few years, much attention has been devoted towards the development of such proof techniques and side-channel 
countermeasures, particularly in the areas of security of finite-length keys 
\cite{Tomamichel2012, tsurumaru, finite, Hayashi2014, Geneva2015} and detector side-channel attacks \cite{mdi,mdi:t1, mdi:t2, mdi:e}.

Amongst these theoretical results, only a few considered the issue of state preparation flaws---despite 
that it is a commonly faced experimental problem. 
More concretely, typical light sources used in QKD systems are not true single-photon sources and practical optical 
modulators employed to encode the light pulses are inherently limited in precision. 
The former can be resolved by using the decoy-state method~\cite{decoy1, decoy2, decoy3}, 
which allows QKD systems based on practical light sources to achieve the security performance of single-photon QKD. 
The latter, however, does not have an adequate solution. 
In particular, it has been firstly shown by Gottesman {\it et al}~\cite{GLLP} that such inaccuracies 
in encoding can lead to very pessimistic secret key rates in the presence of high quantum channel loss. 
Also, other works show similar results~\cite{preskill}. 
This strong dependency on channel loss is primarily due to the fact that state preparation flaws can be seen as a form of basis 
information leakage, which gives Eve some advantage in formulating basis-dependent attacks. 
Crucially, as shown in \cite{GLLP,preskill}, Eve's advantage can be significantly enhanced by 
exploiting channel losses. 
Consequently, this heavily penalizes the secret key rate whenever the channel loss is substantial. 

Very recently, a loss-tolerant QKD protocol~\cite{loss} has been proposed by Tamaki {\it et al} as a means to overcome 
typical encoding flaws in QKD systems. More specifically, as briefly mentioned earlier, here we are considering encoding flaws 
due to imprecise alignment of optical modulators. For example, 
if the quantum states are encoded into the polarisation degree-of-freedom of photons, an encoding flaw 
could be due to a misalignment in the wave-plate used to set the desired polarisation. 
The protocol is similar to the Bennett-Brassard 1984 (BB84) QKD scheme~\cite{qkd1}, 
but instead of considering all the four BB84 states, it uses only three of them. 
Interestingly, by considering statistics beyond those of the BB84 protocol, the resulting secret key rate 
is the same as the one of BB84's~\cite{s1, s2, s3, s8, s6}. 
More importantly, the secret key rate has the very nice property in that it is almost independent of encoding flaws. 
These results imply that the usual stringent demand on precise state preparation can be considerably relaxed 
and one only needs to know the prepared states. 
Additionally, it is useful to mention that most current BB84 QKD systems can easily switch to the loss-tolerant QKD protocol 
without much hardware modifications. 

In anticipation that the loss-tolerant QKD protocol will be widely implemented in the near future, we extend the security 
analysis in Ref.~\cite{loss} to the finite-key regime, {\it i.e.}, we derive explicit bounds on the extractable secret key 
length 
(in \cite{feihu}, the authors have implemented the loss-tolerant protocol experimentally with careful verification of the qubit 
assumption used in the protocol. 
This paper also includes some finite-key analysis of the protocol. 
Unfortunately, however, its phase error rate estimation seems to be valid only against collective attacks). 
Furthermore, our bounds can be applied to a wide range of imperfect light sources---including typical cases whereby the intensity 
of the laser is fluctuating between a certain range. 
Also, the security bounds are obtained within the so-called universal-composable framework~\cite{composable, composable2}, 
and thus secret keys generated using these bounds can be applied to other cryptographic tasks like the one-time-pad. 
In order to investigate the feasibility of our results, we consider a QKD system model that borrows parameters from recent 
fibre-based QKD experiments. With this realistic model, our numerical simulations show that provably-secure keys 
can be distributed up to a fibre length of about 120 km, even when only $10^{11}$ signals are sent by Alice to Bob. 

This paper is organised as follows. In section I\hspace{-.1em}I, we describe some assumptions that 
we made in our security analysis and after that we introduce our protocol. In section I\hspace{-.1em}I\hspace{-.1em}I, 
we give the security definition of the protocol and provide the formulation of the extractable secret key length. 
In section I\hspace{-.1em}V, we present the results of the parameter estimation using the decoy-state method 
for two different cases: an exact intensity control case and an intensity-fluctuation case. 
Then, in section V, we simulate the key generation rate for both scenarios. Finally, section VI concludes the paper with a summary. 
The paper includes as well some Appendixes with additional calculations. 

\section{I\hspace{-.1em}I. assumptions and description of the protocol}
\begin{figure}[t] 
 \begin{center}
 \includegraphics[width=8.5cm,clip]{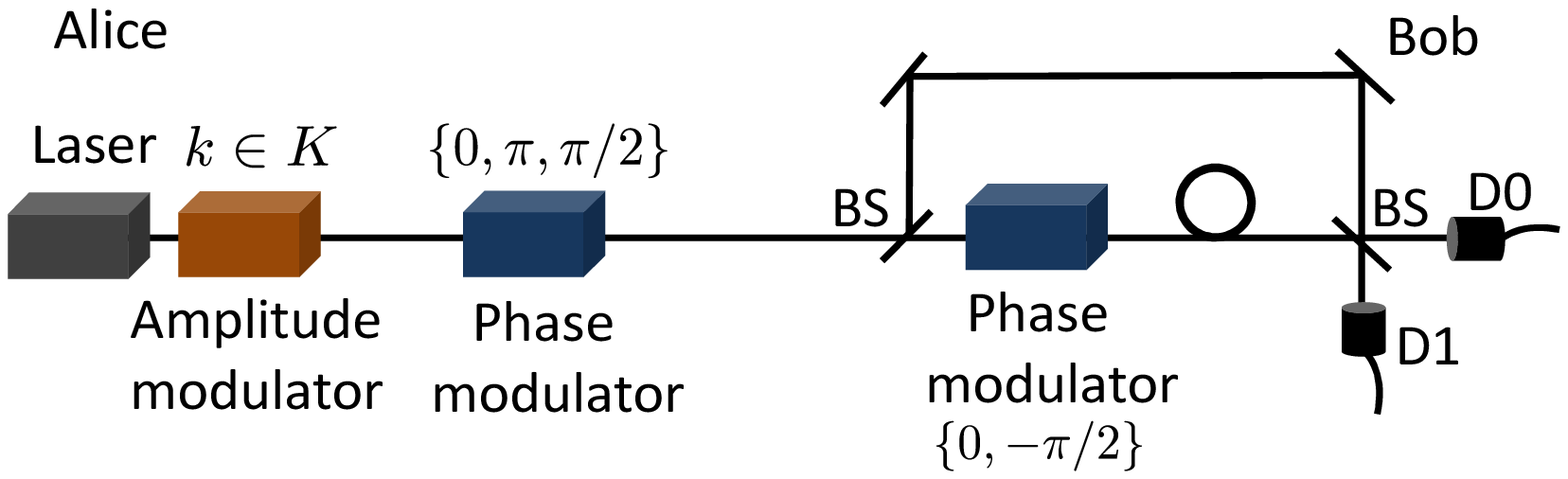}
\end{center}
\caption{
In each trial, Alice's laser emits two consecutive coherent pulses representing the signal and the reference pulse. 
For this, she first uses an amplitude modulator to select the pulses' intensity $k\in K$. 
After that, she applies a phase shift $\{0, \pi,\pi/2\}$ to the signal pulse. 
On reception, Bob splits the received pulses into two beams and then applies a phase shift $\{0,-\pi/2\}$ to one of them. 
Also, he applies a one-pulse delay to one of the arms of the interferometer and then recombine the pulses at a 
50:50 beamsplitter (BS). 
A ``click'' in detector D0 (D1) provides Bob the key bit $y'=0$ ($y'=1$).}
\label{fig:actual}
\end{figure}

\subsection{A. Assumptions on Alice and Bob's devices}
Prior to stating the actual protocol, we first describe the assumptions on the user's devices. 
\newline
\newline
We consider that Alice's transmitter contains a laser source, an amplitude modulator and a phase modulator. 
See Fig. \ref{fig:actual}. 
The laser is single-mode and emits signals with a Poissonian photon number distribution. 
Also, we assume that Alice encodes the bit and the basis information in the relative phase $\theta_{\U{A}}$ between 
a signal and a reference pulse, whose joint phase is perfectly 
randomised
\footnote{Note that the recent work shows that discrete phase randomisation is sufficient for in a BB84 protocol
~\cite{discphase}.}.
Let us emphasise, however, that the security proof that we provide in this paper applies as well to other coding schemes
like, for instance, the polarisation or the time-bin coding schemes. 
Next we present the two types of imperfections that we consider for Alice's device.
\newline
\newline
1. \textit{Intensity fluctuations.}
\newline
The fluctuation of the intensity of the emitted coherent light is typically due to the laser source and imperfections 
in the amplitude modulator. 
Here we shall consider that Alice does not have a full description of the 
probability density function of the fluctuations, but she only knows 
their range \footnote{Note that in those scenarios where Alice knows the exact probability distribution 
of the fluctuations then the conventional decoy-state method can be directly applied.}. 
That is, she knows that the intensity $k$ of the emitted coherent light lies in an interval $k\in [k^-, k^+]$ 
except with error probability $\epsilon_{\U{inten}}$, 
where $k^{+(-)}$ is the upper~(lower) intensity. 
For simplicity, we shall assume that $\epsilon_{\U{inten}}=0$. 
If $\epsilon_{\U{inten}}>0$ this error probability can be directly taken into account through the 
security parameter $\epsilon_{\U{sec}}$ whose definition is referred to equation (\ref{composability}). 
The intensities of the signal and reference pulses are $k^{\U{sig}}:=kV$ 
and $k^{\U{ref}}:=k(1-V)$ respectively, with $0<V<1$. 

In section I\hspace{-.1em}V. A we study the case where $k=k^-=k^+$, {\it i.e.}, there are no intensity fluctuations. 
After that, in section I\hspace{-.1em}V. B, we evaluate the typical scenario where $k^+ > k^-$. 
\newline
\newline
2. \textit{Imperfect encoding of the bit and basis information.}
\newline
In our protocol, Alice chooses the relative phase $\theta_{\U{A}}$ at random from $\{0, \pi/2,\pi\}$ to 
encode the bit and basis information. 
The phase $\theta_\U{A}\in\{0, \pi\}$ corresponds to the $Z$ basis states which are selected with equal probability, 
and $\theta_\U{A}=\pi/2$ denotes the $X$ basis state. Alice assigns a bit value $y=0$ to $\theta_\U{A}\in\{0,\pi/2\}$ 
and a bit value $y=1$ to $\theta_\U{A}=\pi$.

Due to the misalignment of the optical system, however, the actual relative phase prepared by Alice 
may deviate from the desired angle $\theta_{\U{A}}$ by a factor $\Delta\theta_{\U{A}}$. 
Hence, we have that the actual state Alice sends to Bob can be typically described as  
\begin{align}
\int_{0}^{2\pi} p(\Delta\theta_{\U{A}})
P\left[\ket{\sqrt{k^{\U{ref}}}e^{\U{i}\chi}}_{\U{r}}\ket{\sqrt{k^{\U{sig}}}e^{\U{i}(\chi+\theta_{\U{A}}+\Delta\theta_{\U{A}})}}_{\U{s}}\right] 
{\textrm d}\Delta\theta_{\U{A}}.
\label{Astates}
\end{align}
Here, we define $P[\cdot]=\ket{\cdot}\bra{\cdot}$, the parameter $\chi\in[0,2\pi)$ is a random phase, the state 
$\ket{\alpha}_{\U{s(r)}}$ is the coherent state of the signal~(reference) pulse, and 
$p(\Delta\theta_{\U{A}})$ is the probability distribution of $\Delta\theta_{\U{A}}$. 

Alice does not need to know the origin of the encoding errors $\Delta\theta_{\U{A}}$, 
but we assume that she knows $p(\Delta\theta_{\U{A}})$. 
Also, we assume that $p(\Delta\theta_{\U{A}})$ is independently and identically 
distributed for each run of the protocol. 
Moreover, we consider that there are no side-channels in Alice's device. 
\newline
\newline
{\it \textbf{Assumptions on Bob's  apparatus}}\\
We consider that the detection efficiency of Bob's detectors is independent of his measurement basis choice. 
A phase value $\theta_{\U{B}}=0$ ($\theta_{\U{B}}=-\pi/2$) corresponds to 
a device parameter to choose the $Z~(X)$ basis for the measurement. 
Also, like in the case of Alice, we consider that Bob uses an imperfect phase modulator that shifts the phase of the 
incoming signals by $\theta_{\U{B}}+\Delta\theta_{\U{B}}$, where $\Delta\theta_{\U{B}}$ is the modulation error. 
Note, however, that this last assumption is not needed in the security proof; we use it only for simulating the 
resulting secret rate. 
Furthermore, we assume that there are no side-channels in Bob's device. 

\subsection{B. Protocol description}
We study a three-state protocol that uses one signal and two decoy settings. 
Also, we consider that the protocol employs an asymmetric coding, {\it i.e.}, the $Z$ and the $X$ basis are chosen with 
probabilities $p_z$ and $p_x=1-p_z$, respectively. 
The secret key is extracted only from those events where both Alice and Bob select the $Z$ basis and the signal setting. 
In addition, we assume that Alice and Bob do not implement a random sampling procedure to estimate 
the bit error rate, but they perform error correction for a pre-established fixed value of it. 
The error verification step of the protocol (see Step~5 below) informs them about whether or not 
the actual residual bit error rate exceeds the {considered value. %{kon

The protocol runs as follows. 
\newline
\begin{breakbox}
{\it {\textbf{Actual protocol}}}\\\\
First, Alice and Bob decide a security parameter $\epsilon_{\U{sec}}$ whose definition is referred to equation (\ref{composability}). 
Then, they repeat the first three steps of the protocol for $i = 1,...,N$ until the conditions in the Sifting step are met.
\\\\{1. \it Preparation}\\
For each $i$, Alice randomly chooses the intensity 
$k\in K=\{k_{\U{s}}, k_{\U{d1}}, k_{\U{d2}}\}$ with probability
$p_{k_\U{s}}$, $p_{k_{\U{d1}}}$ and $p_{k_{\U{d2}}}=1-p_{k_\U{s}}-p_{k_{\U{d1}}}$, 
respectively. 
The intervals $[k^-,k^+]$ where the different intensities 
lie have to satisfy $k^-_{\U{d1}}>k^+_{\U{d2}}$ and $k^-_{\U{s}}>k^+_{\U{d1}}+k^-_{\U{d2}}$. 
Then, Alice randomly selects the basis $a\in\{Z, X\}$ with probabilities $p_z$ and $p_x$, respectively. 
Next, she chooses at random the signal phase $\theta_{\U{A}}\in\{0, \pi\}$ when she selects the $Z$ basis, 
and she chooses $\theta_{\U{A}}=\pi$ when she selects the $X$ basis. 
Finally, she generates the signal and reference pulses following these specifications and sends 
them to Bob via the quantum channel. 
\\\\{2. \it Measurement}\\
Bob measures the incoming signal and reference pulses using the measurement basis $b\in\{Z, X\}$, 
which he randomly selects with probabilities $p_z$ and $p_x$, respectively. 
The outcome is recorded in $d\in\{0, 1, \perp, \emptyset\}$, where $\perp$ and $\emptyset$ represent the double 
click event and the no click event, respectively. 
If $d=\perp$, Bob assigns a random bit to it \footnotemark. 
As a result, he obtains $y'=\{0,1,\emptyset\}$. 
\\\\ {3. \it Sifting}\\
Alice and Bob announce their bases 
and intensity choices over an authenticated public channel and identify the following sets: 
$Z_k:=\{i|a=b=Z\wedge\U{Intensity}=k\wedge{}y'\neq{}\emptyset\}, X^{j}_k:=\{i|a=b=X\wedge{\U{Intensity}}=k\wedge{}y'=j\}$, 
$Z^{0(1)}X^{j}_k:=\{i|a=Z\wedge{}b=X\wedge{\U{Intensity}}=k\wedge{}y=0~(1)\wedge{}y'=j\}$ and 
$XZ^{j}_k:=\{i|a=X\wedge{}b=Z\wedge{\U{Intensity}}=k\wedge{}y'=j\}$ with $j\in\{0,1\}$ and $k\in{}K$. 
Then, they check if the following conditions are met: $|Z_k|\geq{N_{Z_k}}$, 
$|X^{j}_k|\geq{N_{X^{j}_k}}, |Z^{0(1)}X^{j}_k|\geq{N_{Z^{0(1)}X^{j}_k}}$ and 
$|XZ^{j}_k|\geq{N_{XZ^{j}_k}}$ for all $j\in\{0,1\}$, all $k\in{}K$, and for 
certain pre-established values $N_{Z_k}$, $N_{X^{j}_k}$, $N_{Z^{0(1)}X^{j}_k}$ and $N_{XZ^{j}_k}$, 
where $|\ast|$ represents the length of the set $\ast$. 
\newline
We denote by $N$ the number of pairs of coherent states ({\it i.e.}, signal and reference pulses) sent by Alice 
until these conditions are fulfiled. 
We denote Alice and Bob's sifted keys as $(Z_{{\rm A}}, Z_{{\rm B}})$; 
their size is $|Z_{\rm A}|=|Z_{\rm B}|=|Z_{k_{\U{s}}}|$. 
\\\\ {4. \it Parameter estimation}\\
They estimate the number of events $m_{0(1)}$ where Alice emitted the vacuum (the single-photon) state within the set 
$Z_{k_{\U{s}}}$. 
Their expression is given by equations (\ref{resultm0}) and (\ref{resultm1}) for the scenario without 
intensity fluctuations, and by equations (\ref{resultm0in}) and (\ref{resultm1in}) for the case with intensity fluctuations. 
Also, Alice and Bob estimate $N_{\U{ph}}$, {\it i.e.}, the number of the so-called 
phase errors in the single-photon emissions 
within the set $Z_{k_{\U{s}}}$ (see equation (\ref{Nphase})). 
They check if the phase error rate $e_{\U{ph}}:=N_{\U{ph}}/m_1$ is lower than a predetermined threshold value $\overline{e_{\U{ph}}}$, 
which corresponds to the phase error rate associated with a zero secret key rate (see equation (\ref{keyrate})). 
If $e_{\U{ph}}\geq \overline{e_{\U{ph}}}$ they abort the protocol; otherwise they proceed to step 5.
\\\\{5. \it Postprocessing}\\
Alice and Bob perform error correction over an authenticated public channel for 
$(Z_{{\rm A}}, Z_{{\rm B}})$. 
This step consumes at most $\lambda_{\U{EC}}$ bits. 
Finally, they implement an error verification step and, after that, they perform privacy amplification using 
a hash function that extracts a secret key pair $(S_{{\rm A}}, S_{{\rm B}})$, where $|S_{{\rm A}}|=|S_{{\rm B}}|=\ell$ bits. 
\end{breakbox}
\footnotetext
{Note that this random assignment is not mandatory, and Bob can always choose a particular bit value, say 0, 
for $d=\perp$ as this preserves the basis-independence detection efficiency condition.}

\section{I\hspace{-.1em}I\hspace{-.1em}I. security bounds}
The security of a QKD protocol is characterised by its {\it correctness} and {\it secrecy}.  
That is, following the universal composable security framework~\cite{composable,composable2},  the protocol is called 
$\epsilon_{\U{sec}}$-secure if it is both $\epsilon_{\U{c}}$-correct and $\epsilon_{\U{s}}$-secret, 
where $\epsilon_{\U{sec}}=\epsilon_{\U{c}}+\epsilon_{\U{s}}$. 
Here, the correctness criterion is met whenever the output keys, $S_{\U{A}}$ and $S_{\U{B}}$, are identical. 
More generally, for some small error $\epsilon_\U{c}$  in the correctness, we say that the protocol is 
$\epsilon_\U{c}$-correct if $\Pr[S_{\U{A}} \not= S_{\U{B}}]\leq \epsilon_\U{c}$ is met. 
For the secrecy criterion, it is met whenever the joint classical-quantum state describing Alice's output key 
and Eve's quantum system is of the following form, $U_{\U{A}} \otimes \rho_{\U{E}}$, where $U_{\U{A}}$ is 
the uniform distribution over all bit strings, and $\rho_{\U{E}}$ is an arbitrary quantum state held by Eve. 
Likewise, for some small error $\epsilon_\U{s}$, we say the protocol is $\epsilon_\U{s}$-secret if 
\[  \frac{1}{2}\| \rho_{\U{S}_{\U{A}} \U{E}}-U_{\U{A}}\otimes{}{\rho}_{\U{E}} \|_1 \leq \epsilon_\U{s}, \] where $\rho_{\U{S_A} \U{E}}$ 
is the joint state shared by Alice and Eve. Note that $||\cdot||_1$ is the trace norm defined as 
$||\cdot||_1=\U{Tr}\sqrt{\cdot^\dag\cdot}$. 
Using these security definitions, it can be shown (see Appendix A for details) 
that a lower bound on the secret key length for the protocol described above  

\begin{align}
\ell\geq\Big\lfloor{}m^{\U{L}}_0+m^{\U{L}}_1[1-h(e^{\U{U}}_{\U{ph}})]-\log_2\frac{2}{\epsilon^2_{\U{s}}-\eta}
-\lambda_{\U{EC}}-\log_2\frac{2}{\epsilon_{\U{c}}}\Big\rfloor,
\label{keyrate}
\end{align}
where $h(x)=-x\log_2x-(1-x)\log_2(1-x)$ is the binary entropy function,
$m^{\U{L}}_{0(1)}$ is a lower bound on $m_{0(1)}$, $e^{\U{U}}_{\U{ph}}:=N^{\U{U}}_{\U{ph}}/m^{\U{L}}_1$ 
is an upper bound on the phase error rate, and $\eta$ is the sum of the failure probabilities when estimating $m_{0}$ and $e_{\U{ph}}$. 
This last parameter is upper bounded by $\eta\leq{}1-\U{E}_{Z,0}\U{E}_{Z,1}\U{E}_{\U{ph}}$, where 
$\U{E}_{Z,0}$, $\U{E}_{Z,1}$ and $\U{E}_{\U{ph}}$ are the failure probabilities associated to the estimation of 
$m^{\U{L}}_{0}$, $m^{\U{L}}_{1}$ and to the upper bound on the number of the phase errors $N^{\U{U}}_{\U{ph}}$, 
respectively. 

\section{I\hspace{-.1em}V. parameter estimation}
\label{paraes}
In this section, we briefly describe the estimation procedure to obtain $m^{\U{L}}_{0}$ and $m^{\U{L}}_{1}$. 
Also, we provide an expression for $N_{\U{ph}}^{\U{U}}$. The detailed calculations are included in Appendix D. 

As mentioned in section I\hspace{-.1em}I. B, we assume that the phase of each pulse generated by the laser is 
perfectly randomised. 
This means, in particular, that we can regard the signals sent by Alice as a classical mixture of Fock states, 
each of them representing the total 
number of photons contained both in the signal and in the reference pulse. 
That is, the probability that Alice emits a pulse with $n$ photons conditioned on the fact that she selects the intensity 
setting $k\in K$ is written as 
\begin{align}
p(n|k)= e^{-k}\frac{k^n}{n!}.
\end{align}
Also, from the property of the decoy state method we have that 
the total number of detection events when both Alice and Bob use the $Z$ basis is given by 
\begin{align}
|Z_{\U{tot}}|:=\sum_{k\in{}K}|Z_k|=\sum_{n=0}^{\infty}S_{Z,n},
\label{eq:totalgain}
\end{align}
where $S_{Z,n}$ represents the number of detection events when Alice and Bob used the $Z$ basis and Alice emitted an 
$n$-photon state. 
\subsection{A. Estimation of the number of vacuum and single-photon contributions for the exact intensity control case}
We consider first the scenario without intensity fluctuations in the source, {\it i.e.,} when $k=k^-=k^+$. 

Owing to the use of decoy-states~\cite{decoy1,decoy2,decoy3}, it can be shown that Eve cannot obtain any useful information 
about 
Alice's intensity choice if she observes an $n$-photon state in the quantum channel. 
Therefore, it can be demonstrated that the {\it actual} protocol, where Alice chooses the intensity of each signal 
before she actually sends it to Bob, 
is equivalent to a {\it counterfactual} protocol described as follows. 
First, Alice prepares and sends $n$-photon states to Bob. 
Then, Bob measures all the signals received from Alice. 
Afterwards, Alice decides the intensity setting for each signal. 
Due to this equivalence between the actual and the counterfactual protocols, we have that the number of detection events 
$|Z_{k}|$ for setting $k\in{K}$ within $|Z_{\U{tot}}|$ has the form 
\begin{align}
|Z_{k}|&=\langle{}Z_{k}\rangle+\delta_{k},
\label{eq:gain}
\end{align}
except with certain error probability that will be introduced later on, and where $\langle{}Z_{k}\rangle$ denotes 
the mean value of $|Z_{k}|$ given by 
\begin{align}
\langle{}Z_{k}\rangle&=\sum_{n=0}^{\infty}p(k|n)S_{Z,n}.
\label{eq:gain2}
\end{align}
The parameter $\delta_{k}$ that appears in equation~(\ref{eq:gain}) denotes the deviation between the experimentally obtained quantity 
$|Z_{k}|$ and its expected value. The convergence of $\delta_{k}$ is discussed in Appendix B. 

\subsubsection{a.~~Estimation of the number of vacuum contributions}
At first, we calculate a lower bound on $m_0$, the number of events in $Z_{k_{\U{s}}}$ that originate from a vacuum 
state sent by Alice. 
We define the mean value of $m_0$ as $\mu_0=p(k_{\U{s}}|0)S_{Z,0}$. 
Now, by applying {\it Lemma~1} from Appendix B we obtain that
\begin{align}
m_0\geq\mu_0-\Delta_{Z,0},
\end{align}
except with certain error probability $\epsilon_{Z,0}$, where the 
deviation $\Delta_{Z,0}$ is given by $\Delta_{Z,0}=g_{\U{C}}(\mu_0,\epsilon_{Z,0})$ with ${g}_{\U{C}}(x,y)=\sqrt{2x\ln{1/y}}$. 
So far, the lower bound on $m_{0}$ depends on the unknown mean value $\mu_{0}$ which cannot be directly 
observed in the experiment. 
According to the definition of $\mu_0$, however, this problem can be solved by estimating a lower bound on 
$S_{Z,0}$. 
For this, we use a result from \cite{finite}. 
In particular, we have that 
\begin{align}
S_{Z,0}\geq{}\frac{p(0)}{k_{\U{d1}}-k_{\U{d2}}}
\Big(\frac{k_{\U{d1}}e^{k_{\U{d2}}}}{p_{k_{\U{d2}}}}\langle{}Z_{k_{\U{d2}}}\rangle
-\frac{k_{\U{d2}}e^{k_{\U{d1}}}}{p_{k_{\U{d1}}}}\langle{}Z_{k_{\U{d1}}}\rangle\Big)=:S^{\U{L}}_{Z,0},
\label{vacgain}
\end{align}
where $p(0)=\sum_{k\in{}K}p_k~p(0|k)$. 
To estimate the mean values $\langle{}Z_{k_{\U{d1}}}\rangle$ and $\langle{}Z_{k_{\U{d2}}}\rangle$, 
we can employ either {\it Lemma 2} or {\it Lemma} {\it 3} introduced in Appendix B, such that the fluctuation is minimised. 
In so doing, we obtain a lower bound on $\langle{}Z_{k_{\U{d2}}}\rangle$ together with 
an upper bound on $\langle{}Z_{k_{\U{d1}}}\rangle$ given by
\begin{align}
\langle{Z^-}_{k_{\U{d2}}}\rangle:=|Z_{k_{\U{d2}}}|-\min\Big\{g_{\U{M}}\Big{(}|Z_{k\U{d2}}|,(\epsilon^{k\U{d2}}_{Z,0})^{3/2}
\Big{)}, g_{\U{H}}(|Z_{\U{tot}}|,\epsilon^{k\U{d2}}_{Z,0})\Big\},
\label{decoykd2vac}
\\
\langle{Z^+}_{k_{\U{d1}}}\rangle:=|Z_{k_{\U{d1}}}|+\min\Big\{g_{\U{M}}\Big{(}|Z_{k\U{d1}}|,(\epsilon^{k\U{d1}}_{Z,0})^{4}
/16\Big{)}, g_{\U{H}}(|Z_{\U{tot}}|,\epsilon^{k\U{d1}}_{Z,0})\Big\},
\label{decoykd1vac}
\end{align}
where $g_{\U{M}}(x,y)=\sqrt{2x\ln1/y}$ and $g_{\U{H}}(x,y)=\sqrt{x/2\ln1/y}$.
The failure probability associated with the estimation of $\langle{}Z_{k}\rangle$, with $k\in\{k_{\U{d1}},k_{\U{d2}}\}$, 
is either given by $\varepsilon^{k}_{Z,0}=\epsilon^{k}_{Z,0}$ or by 
$\varepsilon^{k}_{Z,0}=\epsilon^{k}_{Z,0}+\epsilon^{k}_{\U{H},Z,0}$, depending on which 
{\it Lemma} ({\it 2} or {\it 3}) we use. As a result we find that 
\begin{align}
\mu_{0}&\geq p(k_{\U{s}}|0)S^{\U{L}}_{Z,0}
\geq\frac{p_{k_{\U{s}}}e^{-k_{\U{s}}}}{k_{\U{d1}}-k_{\U{d2}}}
\Big(\frac{k_{\U{d1}}e^{k_{\U{d2}}}}{p_{k_{\U{d2}}}}\langle{Z^-}_{k_{\U{d2}}}\rangle-
\frac{k_{\U{d2}}e^{k_{\U{d1}}}}{p_{k_{\U{d1}}}}\langle{Z^+}_{k_{\U{d1}}}\rangle\Big)\nonumber\\
&=:\mu^{\U{L}}_{0}, 
\label{sat}
\end{align}
which only depends on known parameters. Note that in equation~(\ref{sat}) we have used the fact that 
$p(k_{\U{s}}|0)=p_{k_{\U{s}}}p(0|k_{\U{s}})/p(0)=p_{k_{\U{s}}}e^{-k_{\U{s}}}/p(0)$ in combination with equation (\ref{vacgain}). 
We finally obtain, therefore, that 
\begin{align}
m_0\geq{}\mu^{\U{L}}_{0}-\Delta_{Z,0}=:m^{\U{L}}_0,
\label{resultm0}
\end{align}
except with error probability $\varepsilon_{Z,0}=\epsilon_{Z,0}+\varepsilon^{k_{\U{d1}}}_{Z,0}+\varepsilon^{k_{\U{d2}}}_{Z,0}$. 

\subsubsection{b.~~Estimation of the number of single-photon contributions}
Here, we calculate a lower bound on the number of single-photon pulses sent by Alice that contribute to $Z_{k_{\U{s}}}$. 
For this, we use a similar technique to the one described in the previous section. In particular, let $\mu_1$ be the mean value of 
$m_1$, which is given by $\mu_1=p(k_{\U{s}}|1)S_{Z,1}$. 
Then we have that 
\begin{align}
m_1\geq\mu_1-\Delta_{Z,1},
\end{align}
except with error probability $\epsilon_{Z,1}$, where $\Delta_{Z,1}={g}_{\U{C}}(\mu_1,\epsilon_{Z,1})$. 
From Ref.~\cite{finite}, we have that 
\begin{align}
S_{Z,1}&\geq{}\frac{p(1)k_{\U{s}}}{(k_{\U{d1}}-k_{\U{d2}})(k_{\U{s}}-k_{\U{d1}}-k_{\U{d2}})}\Big[
\frac{e^{k_{\U{d1}}}}{p_{k_{\U{d1}}}}\langle{}Z_{k_{\U{d1}}}\rangle
-\frac{e^{k_{\U{d2}}}}{p_{k_{\U{d2}}}}\langle{}Z_{k_{\U{d2}}}\rangle
+\frac{k_{\U{d1}}^2-k_{\U{d2}}^2}{k_{\U{s}}^2}
\Big(\frac{S^{\U{L}}_{Z,0}}{p(0)}-\frac{e^{k_{\U{s}}}\langle{}Z_{k_s}\rangle}{p_{k_{\U{s}}}}\Big)
\Big] \nonumber\\
&=:S^{\U{L}}_{Z,1},
\end{align}
where $p(1)=\sum_{k\in{}K}p_k~p(1|k)$. 
As before, by using {\it Lemmas} 2 and 3 from Appendix B we obtain a lower bound on $\langle{}Z_{k_{\U{d1}}}\rangle$, and 
an upper bound on $\langle{}Z_{k_{\U{d2}}}\rangle$ and $\langle{}Z_{k_{\U{s}}}\rangle$. 
They are given by
\begin{align}
\langle{Z^-}_{k_{\U{d1}}}\rangle:=|Z_{k_{\U{d1}}}|-\min\Big\{g_{\U{M}}\Big{(}|Z_{k\U{d1}}|,(\epsilon^{k\U{d1}}_{Z,1})^{3/2}\Big{)}, 
g_{\U{H}}(|Z_{\U{tot}}|,\epsilon^{k\U{d1}}_{Z,1})\Big\},
\label{decoykd1sin}
\\
\langle{Z^+}_{k}\rangle:=|Z_{k}|+\min\Big\{g_{\U{M}}\Big{(}|Z_{k}|,(\epsilon^{k}_{Z,1})^{4}/16\Big{)}, 
g_{\U{H}}(|Z_{\U{tot}}|,\epsilon^{k}_{Z,1})\Big\},
\label{decoyksin}
\end{align}
where the second equality holds for $k\in\{k_{\U{s}}, k_{\U{d2}}\}$. 
The failure probability associated with the estimation of $\langle{}Z_{k}\rangle$~(with $k\in{K}$) 
is either given by $\varepsilon^{k}_{Z,1}=\epsilon^{k}_{Z,1}$ or by 
$\varepsilon^{k}_{Z,1}=\epsilon^{k}_{Z,1}+\epsilon^{k}_{\U{H},Z,1}$, depending again on which {\it Lemma} ({\it 2} or {\it 3}) 
we apply. 
By employing the relation $p(k_{\U{s}}|1)=p_{k_{\U{s}}}p(1|k_{\U{s}})/p(1)=p_{k_{\U{s}}}k_{\U{s}}e^{-k_{\U{s}}}/p(1)$, 
we obtain a lower bound on $\mu_1$, which only depends on known parameters,
\begin{align}
\mu_1&\geq p(k_{\U{s}}|1)S^{\U{L}}_{Z,1}\nonumber\\
&\geq\frac{p_{k_{\U{s}}}k_{\U{s}}^2e^{-k_{\U{s}}}}{(k_{\U{d1}}-k_{\U{d2}})(k_{\U{s}}-k_{\U{d1}}-k_{\U{d2}})}\Big[
\frac{e^{k_{\U{d1}}}}{p_{k_{\U{d1}}}}\langle{Z}^-_{k_{\U{d1}}}\rangle
-\frac{e^{k_{\U{d2}}}}{p_{k_{\U{d2}}}}\langle{Z}^+_{k_{\U{d2}}}\rangle
+\frac{k_{\U{d1}}^2-k_{\U{d2}}^2}{k_{\U{s}}^2}\Big(\frac{\mu^{\U{L}}_{0}}{p(k_{\U{s}}\wedge 0)}
-\frac{e^{k_{\U{s}}}\langle{Z}^+_{k_{\U{s}}}\rangle}{p_{k_{\U{s}}}}\Big)\Big]\nonumber\\
&=:\mu^L_{1}.
\end{align}
Therefore, we have that
\begin{align}
m_1\geq{}\mu^{\U{L}}_{1}-\Delta_{Z,1}=:m^{\U{L}}_1,
\label{resultm1}
\end{align}
except with error probability 
$\varepsilon_{Z,1}=\sum^1_{n=0}(\varepsilon^{k_{\U{d1}}}_{Z,n}+\varepsilon^{k_{\U{d2}}}_{Z,n})+\epsilon_{Z,1}+\varepsilon^{k_{\U{s}}}_{Z,1}$, 
where the parameters $\varepsilon^{k_{\U{d1}}}_{Z,0}$ and $\varepsilon^{k_{\U{d2}}}_{Z,0}$ come from the estimation of 
$\mu^{\U{L}}_0$. 

\subsection{B. Estimation of the number of vacuum and single-photon contributions for the intensity-fluctuation case}
We now evaluate the scenario where the laser suffers from intensity fluctuations. 
As introduced above, here we shall assume that Alice only knows the range $[k^-,k^+]$ where the intensity value $k$ lies. 
Below we introduce the final expressions for the different parameters; the detailed derivations are referred to Appendix C. 
\subsubsection{a.~~Estimation of the number of vacuum contributions}
Here, we present the result for the estimation of the lower bound on $T_{Z,0}$. 
Here, $T_{Z,0}$ is the sum of the conditional probability that Bob detects a signal in the $Z$ basis conditioned 
that Alice chooses the signal intensity and sends a vacuum state in the $Z$ basis 
(see equation (\ref{TZ0})). 
It is given by
\begin{align}
T_{Z,0}\geq\frac{1}{k^-_{\U{d1}}-k^+_{\U{d2}}}\Big(
\frac{k^-_{\U{d1}}e^{k^-_{\U{d2}}}}{p_{k_{\U{d2}}}}\langle{Z_{k_{\U{d2}}}}\rangle
-\frac{k^+_{\U{d2}}e^{k^+_{\U{d1}}}}{p_{k_{\U{d1}}}}\langle{Z_{k_{\U{d1}}}}\rangle\Big)=:T^{\U{L}}_{Z,0}.
\end{align}
To calculate the mean values $\langle{Z_{k_{\U{d1}}}}\rangle$ and $\langle{Z_{k_{\U{d2}}}}\rangle$ 
we employ Azuma's inequality, which is described in {\it Lemma~4} (see Appendix~B). 
Importantly, note that this inequality holds 
without assuming independence of the trials. 
As a result, we obtain a lower bound on $\langle{Z_{k_{\U{d1}}}}\rangle$ 
together with an upper bound on $\langle{Z_{k_{\U{d2}}}}\rangle$. 
They are given by 
\begin{align}
\langle{Z^-}_{k_{\U{d2}}}\rangle:=|Z_{k_{\U{d2}}}|-g_{\U{A}}(N_{z}, \epsilon^{k\U{d2}}_{Z,0}),
\label{azumanumber1}\\
\langle{Z^+}_{k_{\U{d1}}}\rangle:=|Z_{k_{\U{d1}}}|+g_{\U{A}}(N_z, \epsilon^{k\U{d1}}_{Z,0}),
\label{azumanumber2}
\end{align}
where $g_{\U{A}}(x,y)=\sqrt{2x\ln(1/y)}$, and $N_z$ is the number of events where Alice and Bob use the $Z$ basis 
within $N$ trials. 

In so doing, we find a lower bound on $\mu_0$ that only depends on parameters that are directly observed in the experiment. 
It has the form 
\begin{align}
\mu_0
&\geq p^-(k_{\U{s}}\wedge0)T^{\U{L}}_{Z,0}\nonumber\\
&\geq \frac{p_{k_{\U{s}}}e^{-k^+_{\U{s}}}}{k^-_{\U{d1}}-k^+_{\U{d2}}}
\Big(\frac{k^-_{\U{d1}}e^{k^-_{\U{d2}}}}{p_{k_{\U{d2}}}}\langle{Z^-_{k_{\U{d2}}}}\rangle
-\frac{k^+_{\U{d2}}e^{k^+_{\U{d1}}}}{p_{k_{\U{d1}}}}\langle{Z^+_{k_{\U{d1}}}}\rangle\Big)\nonumber\\
&=:\mu^{\U{L}}_{0},
\end{align}
where the $p^-(k_{\U{s}}\wedge0)$ is a lower bound on $p(k_{\U{s}}\wedge0)$.
\newline
Finally, we obtain a lower bound on $m_0$ which is given by 
\begin{align}
m_0\geq{}\mu^{\U{L}}_{0}-\Delta_{Z,0}=:m^{\U{L}}_0,
\label{resultm0in}
\end{align}
except with error probability $\varepsilon_{Z,0}=\epsilon_{Z,0}+\epsilon^{k_{\U{d1}}}_{Z,0}+\epsilon^{k_{\U{d2}}}_{Z,0}$.

\subsubsection{b. Estimation of the number of single-photon contributions}
Here, we introduce a lower bound on $T_{Z,1}$. 
Here, $T_{Z,1}$ is the sum of the conditional probability that Bob detects a signal in the $Z$ basis conditioned that Alice chooses the 
signal intensity and sends a single-photon state in the $Z$ basis (see equation (\ref{TZ1})). 

It is given by
\begin{align}
T_{Z,1}&\geq\frac{k^-_{\U{s}}}{(k^+_{\U{d1}}-k^-_{\U{d2}})(k^-_{\U{s}}-k^+_{\U{d1}}-k^-_{\U{d2}})}
\Big[\frac{e^{k^-_{\U{d1}}}}{p_{k_{\U{d1}}}}\langle{Z_{k_{\U{d1}}}}\rangle-\frac{e^{k^+_{\U{d2}}}}{p_{k_{\U{d2}}}}
\langle{Z_{k_{\U{d2}}}}\rangle
-\frac{(k^+_{\U{d1}})^2-(k^-_{\U{d2}})^2}{(k^-_{\U{s}})^2}
\Big(\frac{e^{k^+_{\U{s}}}}{p_{k_{\U{s}}}}\langle{Z_{k_s}}\rangle-T^{\U{L}}_{Z,0}\Big)\Big]\nonumber\\
&=:T^{\U{L}}_{Z,1}.
\end{align}
Again, to estimate the mean values $\langle{Z_{k_{\U{d1}}}}\rangle, \langle{Z_{k_{\U{d2}}}}\rangle$ and 
$\langle{Z_{k_{\U{s}}}}\rangle$ we employ {\it Lemma~4}. This way we obtain a lower bound on $\langle{Z_{k_{\U{d1}}}}\rangle$ 
and an upper bound on $\langle{Z_{k_{\U{d2}}}}\rangle$ and $\langle{Z_{k_{\U{s}}}}\rangle$ as
\begin{align}
\langle{Z^-}_{k_{\U{d1}}}\rangle:=|Z_{k_{\U{d1}}}|-g_{\U{A}}(N_z, \epsilon^{k\U{d1}}_{Z,1}),\\
\langle{Z^+}_k\rangle:=|Z_k|+g_{\U{A}}(N_z, \epsilon^{k}_{Z,1}),
\end{align}
where the second equality holds for $k\in\{k_{\U{s}}, k_{\U{d2}}\}$.

Hence, a lower bound on $\mu_1$ can be directly written as 
\begin{align}
\mu_1&\geq p^-(k_{\U{s}}\wedge1)T^{\U{L}}_{Z,1}\nonumber\\
&\geq \frac{p_{k_{\U{s}}}e^{-k^-_{\U{s}}}(k^-_{\U{s}})^2}
{(k^+_{\U{d1}}-k^-_{\U{d2}})(k^-_{\U{s}}-k^+_{\U{d1}}-k^-_{\U{d2}})}
\Big[\frac{e^{k^-_{\U{d1}}}}{p_{k_{\U{d1}}}}\langle{Z^-_{k_{\U{d1}}}}\rangle
-\frac{e^{k^+_{\U{d2}}}}{p_{k_{\U{d2}}}}\langle{Z^+_{k_{\U{d2}}}}\rangle
-\frac{(k^+_{\U{d1}})^2-(k^-_{\U{d2}})^2}{(k^-_{\U{s}})^2}
\Big(\frac{e^{k^+_{\U{s}}}}{p_{k_{\U{s}}}}\langle{Z^+_{k_{\U{s}}}}\rangle
-\frac{\mu^{\U{L}}_{0}}{p^-(k_{\U{s}}\wedge 0)}\Big)\Big]\nonumber\\
&=: \mu^{\U{L}}_{1},
\end{align}
where $p^-(k_{\U{s}}\wedge 1)$ is a lower bound on $p(k_{\U{s}}\wedge1)$.
\newline
Finally, we obtain $m^{\U{L}}_1$ as 
\begin{align}
m_1\geq{}\mu^{\U{L}}_{1}-\Delta_{Z,1}=:m^{\U{L}}_1,
\label{resultm1in}
\end{align}
except with error probability 
$\varepsilon_{Z,1}=\sum^1_{n=0}(\epsilon^{k_{\U{d1}}}_{Z,n}+\epsilon^{k_{\U{d2}}}_{Z,n})+\epsilon_{Z,1}+\epsilon^{k_{\U{s}}}_{Z,1}$.

\subsection{C. Estimation of the number of phase errors}
In this section we present an upper bound on $N_{\U{ph}}$, which is the number of phase errors in the single-photon emissions 
within the set $Z_{k_{\U{s}}}$. 
As already mentioned in section~I\hspace{-.1em}I. B, the states sent by Alice are given by equation~(\ref{Astates}), 
and we assume that the distribution $p(\Delta\theta_{\U{A}})$ is known to Alice. 

We denote the single-photon part of equation (\ref{Astates}) as $\rho(\theta_{\U{A}})$; it is given by
\begin{align}
\rho(\theta_{\U{A}})=
\int^{2\pi}_0 p(\Delta\theta_{\U{A}})\frac{1}{2}
\Big[\sigma_I+\frac{2\gamma}{1+\gamma^2}\Big(\cos(\theta_{\U{A}}+\Delta\theta_{\U{A}})\sigma_Z
+\sin(\theta_{\U{A}}+\Delta\theta_{\U{A}})\sigma_X\Big)+\frac{1-\gamma^2}{1+\gamma^2}\sigma_Y\Big]
\U{d}\Delta\theta_{\U{A}},
\end{align}
where the parameter $\gamma=\sqrt{k^{\U{sig}}/k^{\U{ref}}}$. 
Here we define the eigenvectors of the Pauli operators $\sigma_Y, \sigma_Z$ and $\sigma_X$ as: 
$\ket{0_y}=\ket{1}_{\U{r}}\ket{0}_{\U{s}}, \ket{1_y}=\ket{0}_{\U{r}}\ket{1}_{\U{s}}$, 
$\ket{0_z}=(\ket{0_y}+\ket{1_y})/\sqrt{2}, \ket{1_z}=(-i\ket{0_y}+i\ket{1_y})/\sqrt{2}$ and 
$\ket{i_x}=(\ket{0_z}+(-1)^i\ket{1_z})/\sqrt{2}$ with $i\in\{0,1\}$. 
The state $\ket{n}_{\U{r(s)}}$ denotes an $n$-photon number state of the reference (signal) pulse. 

With this notation, the single-photon part of the three states sent by Alice can be expressed as 
${\rho}_{0z}=\rho(0)$, ${\rho}_{1z}=\rho(\pi)$ and ${\rho}_{0x}=\rho(\pi/2)$. 
Let ${\rho}_{S}=({\sigma}_I+\vec{\sigma}\cdot{\vec{V}_{S}})/2$, where 
$\vec{\sigma}=[{\sigma}_X, {\sigma}_Y, {\sigma}_Z]$ and the Bloch vector 
$\vec{V}_{S}=[V^{S}_X,V_Y,V^{S}_Z]$ is a real three-dimensional vector that satisfies 
$|\vec{V}_S|\leq{1}$ with $S\in\{0z,1z,0x\}$. 
From \cite{loss} we have that if $V_Y\neq{}0$, the phase error rate of $\rho_{0z}$ and $\rho_{1z}$ 
is equivalent to that obtained after the application of the following filter operation, 
\begin{align}
{F}_{Y}=(1+\sqrt{1-V_Y}/\sqrt{1+V_Y}){P}[\ket{0_{y}}]+2V_Y/(-1+V_Y+\sqrt{1-V_Y^2}){P}[\ket{1_{y}}].
 \end{align}
This means, in particular, that we can restrict ourselves to the estimation of the phase error rate of the states 
$\tilde{\rho}_{S}$ which lie in the $\sigma_X$-$\sigma_Z$ plane,
\begin{align}
\tilde{\rho}_{S}:=\frac{{F}_Y{\rho}_S{F}^{\dag}_{Y}}{\U{Tr}[{{F}^{\dag}_Y{F}_Y{\rho}_S}]}=
\frac{(\sigma_I+r^S_x\sigma_X+r^S_z{\sigma}_Z)}{2},
\label{filteredstate}
\end{align}
where the parameters $r^S_x$ and $r^S_z$ are given by $r^S_x=V^S_{X}f(V_Y)$ and $r^S_{z}=V^S_{Z}f(V_Y)$ with $f(V_Y)=1/\sqrt{1-V_Y^2}$. 
The states $\tilde{\rho}_S$ given by equation~(\ref{filteredstate}) can also be decomposed as 
\begin{align}
\tilde{\rho}_S=P^S_{0}P[\ket{\phi^{S}_{0}}]+P^S_{1}P[\ket{\phi^{S}_{1}}],
\label{tilderho}
\end{align}
where the probabilities $P^S_{i}$ have the form 
\begin{align}
P^S_{i}=\frac{1}{2}\Big(1-(-1)^{i}\sqrt{({r^{S}_x})^2+({r^{S}_z})^2}\Big),
\end{align}
and the eigenvectors $\ket{\phi^{S}_{i}}$ are given by
\begin{align}
\ket{\phi^{S}_{i}}&=\begin{cases}
    \frac{1}{\U{\mathcal{N}}}\Big(\frac{r^S_z-(-1)^i\sqrt{(r^S_x)^2+(r^S_z)^2}}{r^S_x}\ket{0_z}+\ket{1_z}\Big) & (r_x\neq0) \\
    \ket{i_z} & (r_x=0 \wedge {}r_z<0)\\
    \ket{i\oplus{1}_z} & (r_x=0 \wedge {}r_z>0)
  \end{cases}\\
&=:a^{S}_{i}\ket{0_z}+b^{S}_{i}\ket{1_z},
\label{eigenvec}
\end{align}
for $i\in\{0,1\}$, and where $\mathcal{N}$ is the normalisation factor of the state. 

After some lengthy calculations (see Appendix D for details), 
we obtain that $N_{\U{ph}}$ is upper bounded by
\begin{align}
N_{\U{ph}}&\leq{}\sum^1_{s=0}
\frac{P(s+1)}{2\{1+(-1)^s(\sqrt{P^{0z}_{0}P^{1z}_{0}}\expect{\phi^{0z}_0|\phi^{1z}_0}+
\sqrt{P^{0z}_{1}P^{1z}_{1}}\expect{\phi^{0z}_1|\phi^{1z}_1})\}}\Big[
N_{M_{Xs}}(3)+N_{M_{Xs}}(4)\nonumber\\
&+(-1)^s\sum^1_{t=0}\sqrt{P^{0z}_{t}P^{1z}_{t}}\Big\{C_{t,0}N_{M_{Xs}}(3)+C_{t,1}N_{M_{Xs}}(4)+C_{t,2}N_{M_{Xs}}(5)\Big\}
\Big]+\Delta^{s\oplus 1}_{\U{A},s+1}\nonumber\\
&=:N^{\U{U}}_{\U{ph}},
\label{Nphase}
\end{align}
except with error probability $\varepsilon_{\U{ph}}$. 
Here, the terms $N_{M_{{X}s}}(j)$ with $j\in\{3,4,5\}$ are defined in equations (\ref{azuma3})-(\ref{azuma5}); the quantities 
$P(1)$ and $P(2)$ are given by equation (\ref{Prob}); the parameters $C_{t,l}$ have the form 
$C_{t,l}:=(a^{0z}_ta^{1z}_t+b^{0z}_tb^{1z}_t)A^{-1}_{0,l}+(a^{0z}_tb^{1z}_t+b^{0z}_ta^{1z}_t)A^{-1}_{1,l}
+(a^{0z}_ta^{1z}_t-b^{0z}_tb^{1z}_t)A^{-1}_{2,l}$ for $l\in\{0,1,2\}$; the coefficients 
$A^{-1}_{i,j}$ are the $(i,j)$ element of the following matrix 
\begin{align}
A^{-1}&:=\frac{1}{Q}
\begin{pmatrix}
2(r_x^{1z}r_z^{0x}-r_x^{0x}r_z^{1z}) & 2(r_x^{0x}r_z^{0z}-r_x^{0z}r_z^{0x}) & 
2(r_x^{0z}r_z^{1z}-r_x^{1z}r_z^{0z})\\
2(r_z^{1z}-r_z^{0x}) & 2(r_z^{0x}-r_z^{0z}) & 2(r_z^{0z}-r_z^{1z})\\
2(r_x^{0x}-r_x^{1z}) & 2(r_x^{0z}-r_x^{0x}) & 2(r_x^{0z}-r_x^{0x})
\end{pmatrix},
\label{matrixA}
\end{align}
where $Q:=r_x^{1z}(r_z^{0x}-r_z^{0z})+r_x^{0x}(r_z^{0z}-r_z^{1z})+r_x^{0z}(r_z^{1z}-r_z^{0x})$; and 
the fluctuation term $\Delta^{s\oplus 1}_{\U{A},s+1}$ is given by equation (\ref{three}). 
 
\section{V. simulation of the key rate}
\begin{figure}[t] 
 \begin{center}
 \includegraphics[width=7cm,clip]{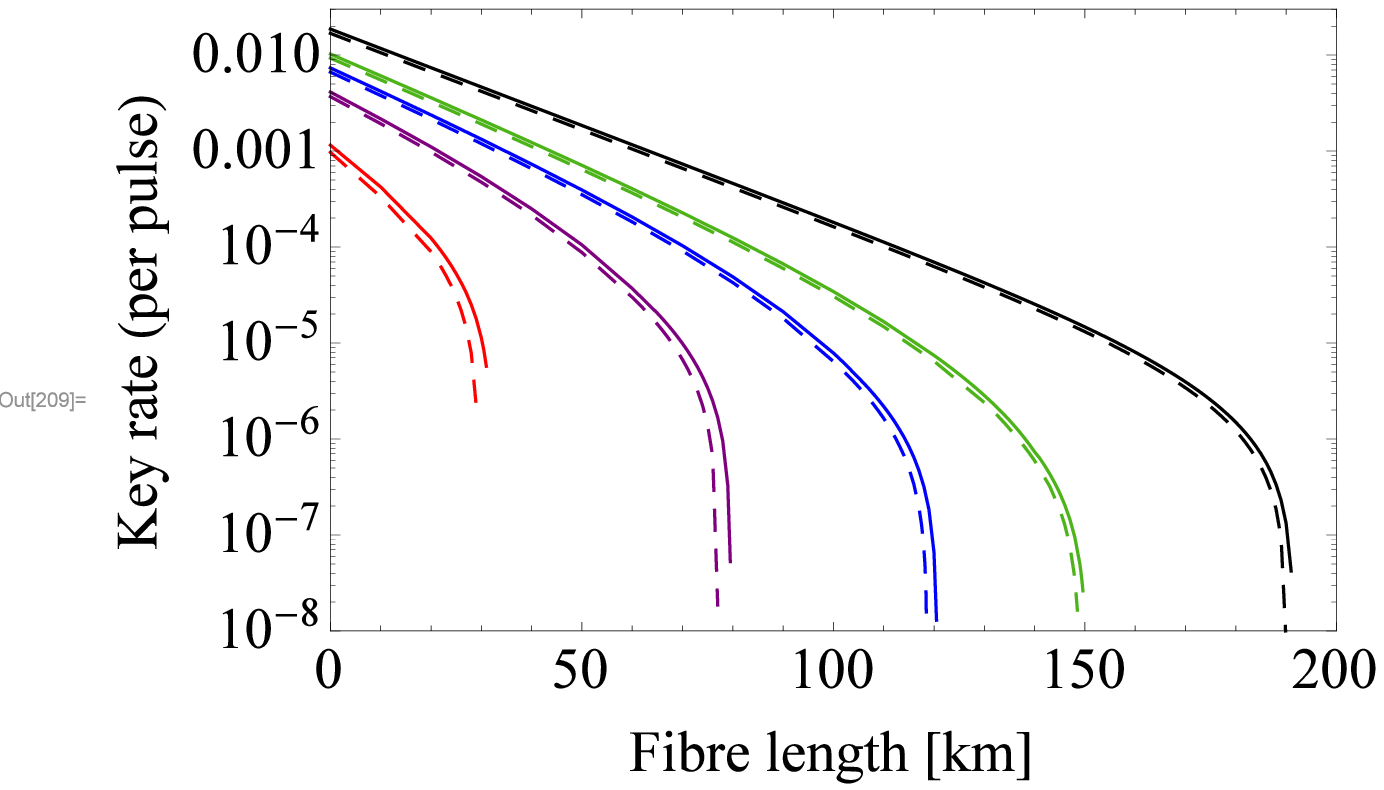}
\end{center}
\caption{(Colour online) Secret key rate (per pulse) in logarithmic scale vs fibre length for the case with exact 
intensity control. The security parameter is $\epsilon_{\U{sec}}=10^{-10}$ and the total number of signals sent by Alice is $N=10^s$ with 
$s=9,10, 11$ and $12$ (from left to right). 
The rightmost two lines correspond to the asymptotic secret key rate with two decoy settings. 
The solid lines denote the case $\xi$=0 ({\it i.e.}, the perfect encoding scenario) while the dashed lines show the case 
$\xi$=0.147 which is equivalent to a phase modulation error of 8.42$^\circ$ 
(this error parameter is measured in an updated version of a commercial plug\&play system (ID Quantique Clavis2)~\cite{feihu}). 
The experimental parameters are described in the main text.
}
\label{fig:keyrate1}
\end{figure}
\begin{figure}[t] 
 \begin{center}
 \includegraphics[width=7cm,clip]{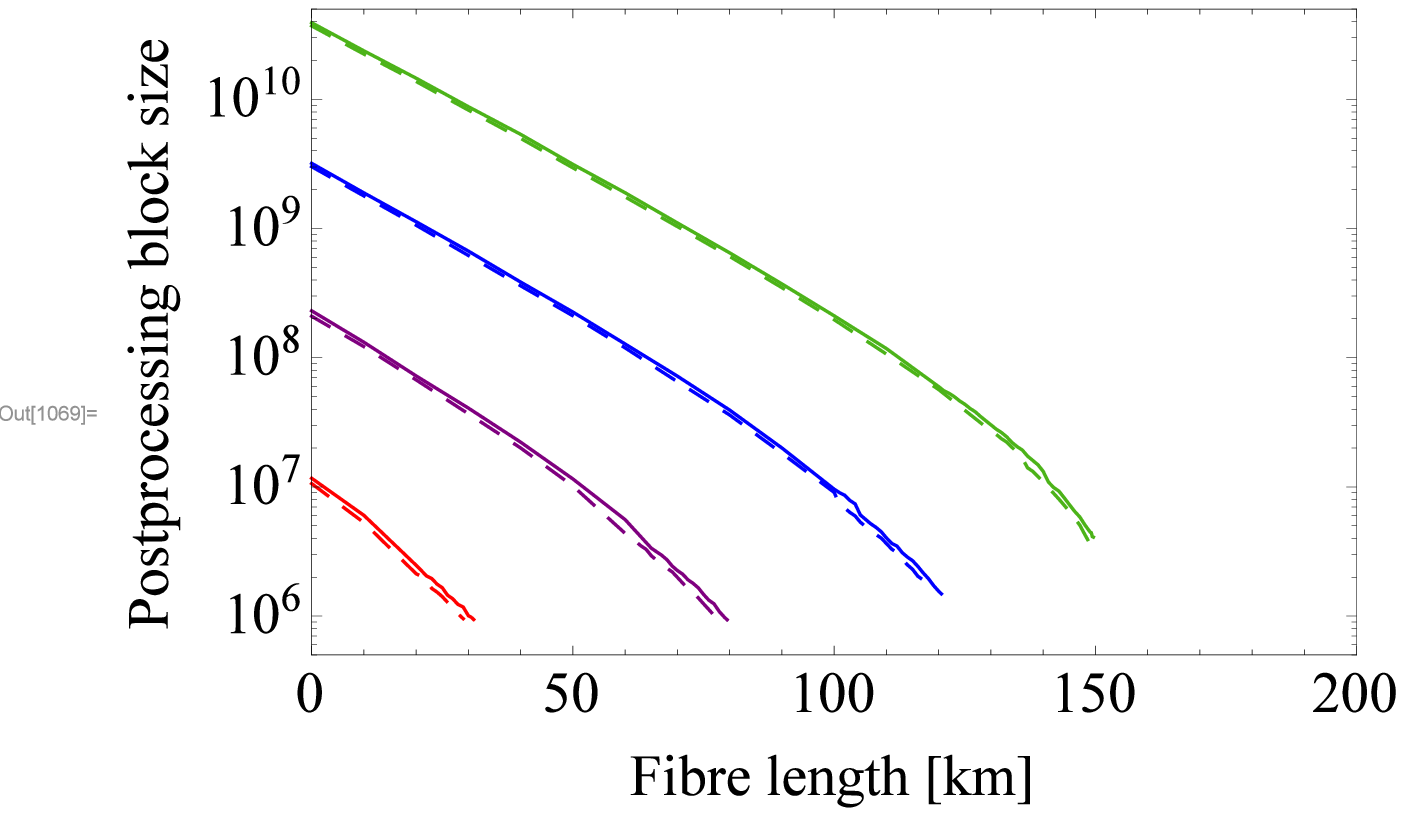}
\end{center}
\caption{(Colour online) Postprocessing block size $|Z_{k_{\U{s}}}|$ vs fibre length for a fixed total number of signals $N=10^s$ 
sent by Alice with exact intensity control, with $s=9,10, 11$ and $12$ (from left to right). 
The solid lines correspond to the case $\xi$=0 and the dashed lines are for $\xi$=0.147.}
\label{fig:block}
\end{figure}
\begin{figure}[t] 
 \begin{center}
 \includegraphics[width=7cm,clip]{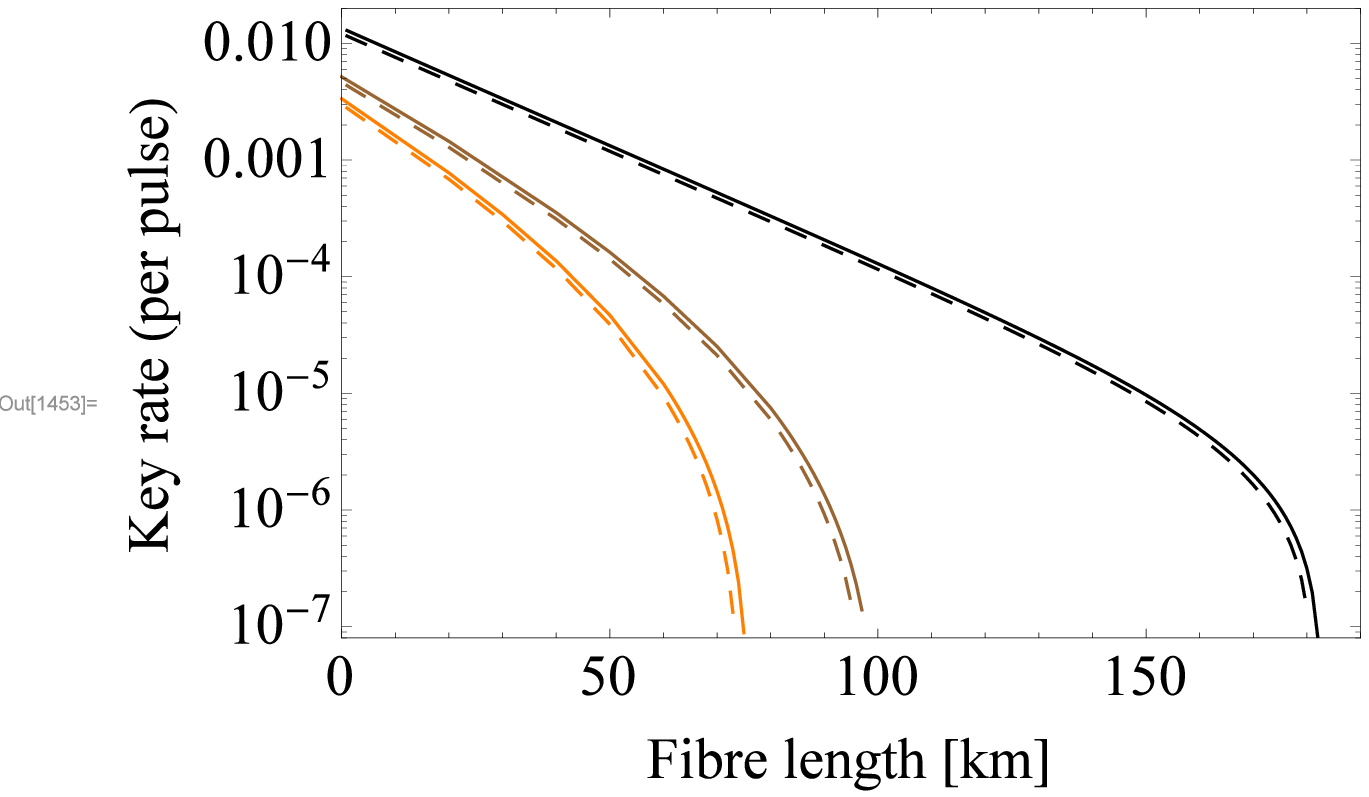}
\end{center}
\caption{(Colour online) Secret key rate (per pulse) in logarithmic scale vs fibre length when the intensity fluctuation is 2$\%$. 
The security parameter is $\epsilon_{\U{sec}}=10^{-10}$ and the total number of signals sent by Alice is 
$N=10^s$ with $s=\{14, 15\}$ (from left to right). The rightmost two lines correspond to the asymptotic secret key 
rate with two decoy settings. 
The solid lines denote the case $\xi=0$ ({\it i.e.}, the perfect encoding scenario) while the dashed lines show the case 
$\xi=0.147$ (which is equivalent to a phase modulation error of 8.42$^\circ$). The experimental parameters are described 
in the main text. }
\label{fig:keyrate2}
\end{figure}
\begin{figure}[t] 
 \begin{center}
   \includegraphics[width=7cm,clip]{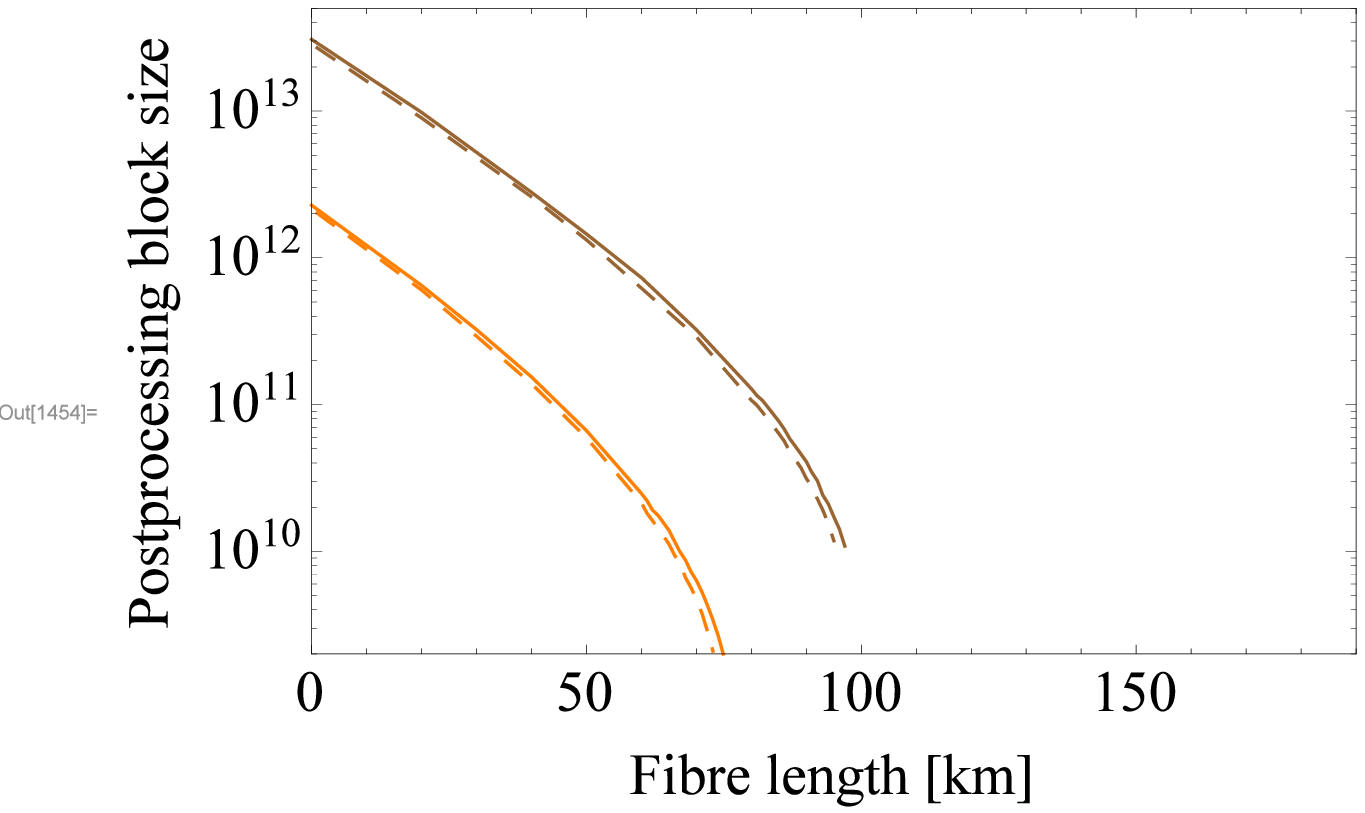}
\end{center}
\caption{
(Colour online) 
Postprocessing block size $|Z_{k_{\U{s}}}|$ vs fibre length for a fixed total number of signals $N=10^s$ 
sent by Alice, with $s=\{14,15\}$ when the intensity fluctuation is 2\%~(from left to right). 
The solid lines correspond to the case $\xi$=0 and the dashed lines are for $\xi$=0.147.
}
\label{fig:PPB1}
\end{figure}
\begin{figure}[t] 
 \begin{center}
   \includegraphics[width=7cm,clip]{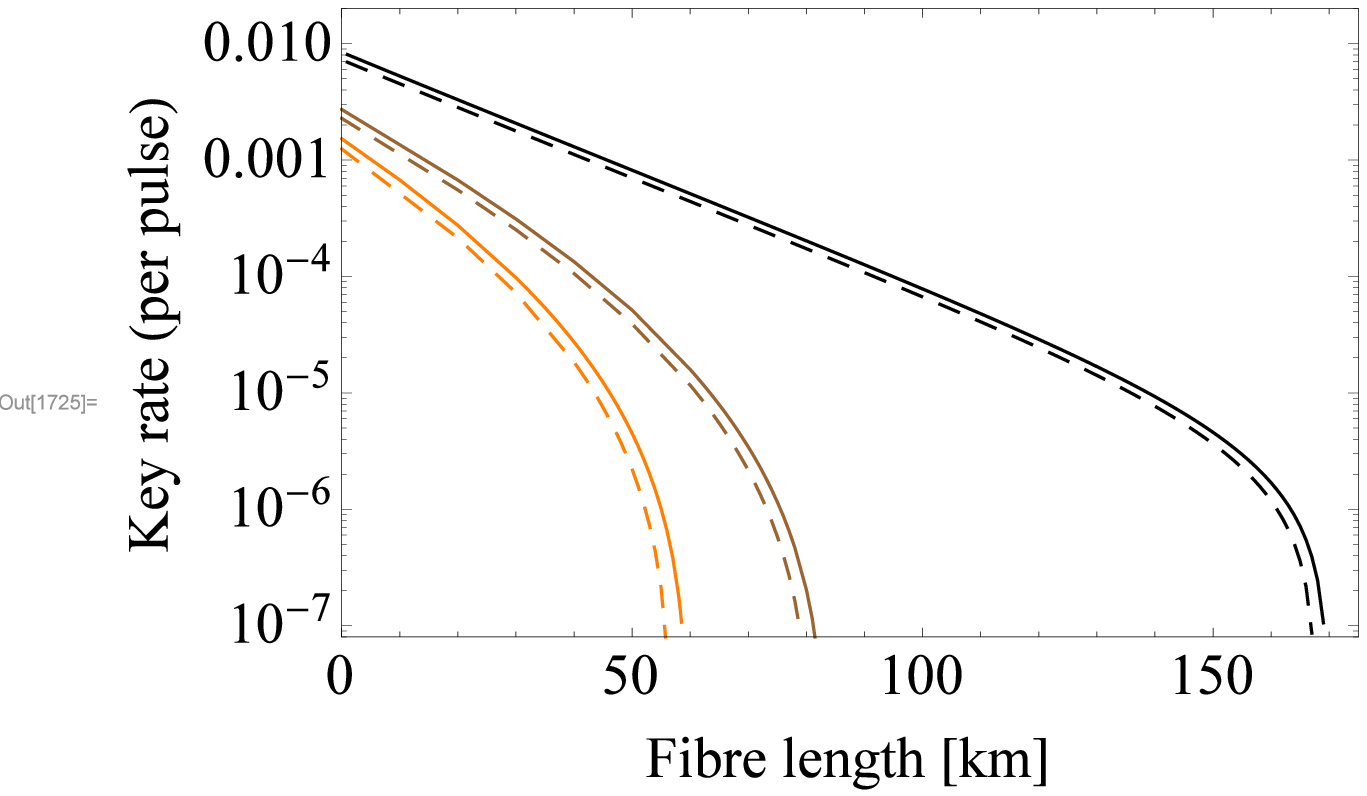}
\end{center}
\caption{(Colour online) Secret key rate (per pulse) in logarithmic scale vs fibre length when the intensity fluctuation is 5$\%$. 
The security parameter is 
$\epsilon_{\U{sec}}=10^{-8}$ and the total number of signals sent by Alice is $N=10^s$ with $s=\{14, 15\}$ (from left to right). 
The rightmost two lines correspond to the asymptotic secret key rate with two decoy settings. 
The solid lines denote the case $\xi=0$ ({\it i.e.}, the perfect encoding scenario) while the dashed lines show the case 
$\xi=0.147$ (which is equivalent to a phase modulation error of 8.42$^\circ$). 
The experimental parameters are described in the main text.
}
\label{fig:keyrate3}
\end{figure}
\begin{figure}[t] 
 \begin{center}
   \includegraphics[width=7cm,clip]{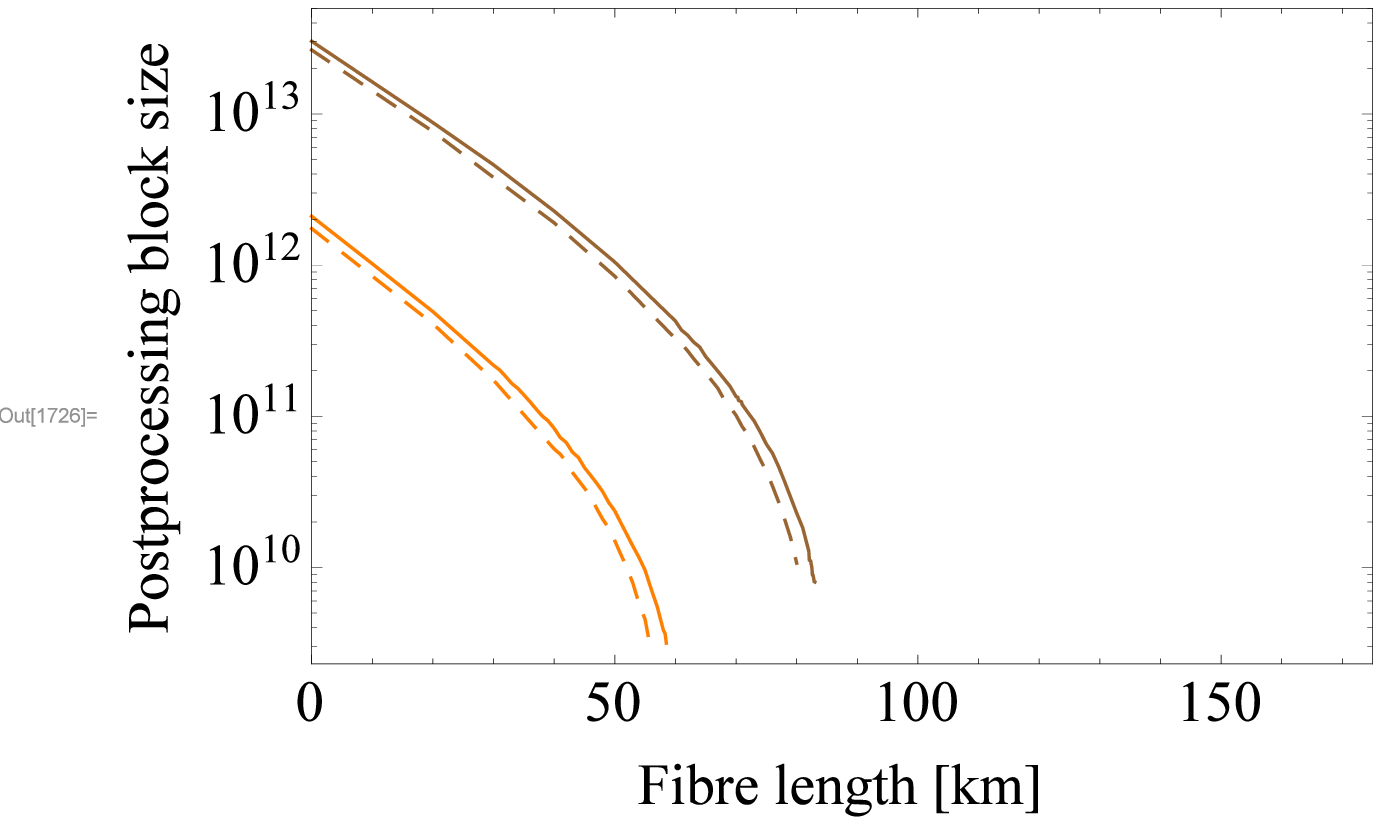}
\end{center}
\caption{
(Colour online) 
Postprocessing block size $|Z_{k_{\U{s}}}|$ vs fibre length for a fixed total number of signals $N=10^s$ 
sent by Alice, with $s=\{14,15\}$ when the intensity fluctuation is 5\% (from left to right). 
The solid lines correspond to the case $\xi$=0 and the dashed lines are for $\xi$=0.147.
}
\label{fig:PPB2}
\end{figure}

In this section, we show the simulation result for a fibre-based QKD system. 
Alice chooses the intensity of the laser from the set $\{k_{\U{s}}, k_{\U{d1}}, k_{\U{d2}}\}$, 
where we fix the intensity of the weakest decoy state to $k_{\U{d2}}=2\times{}10^{-4}$. 
This is so because, in practice, it is difficult to generate a vacuum state due to the imperfect extinction of 
the amplitude modulator. 
Also, we assume that Bob uses an active measurement setup with two single-photon detectors with detection 
efficiency $\eta_{\U{det}}=15\%$ and a dark count probability $p_\U{d}=5\times{}10^{-7}$. The attenuation coefficient of the 
optical fibre is $0.2\U{dB/km}$ and its transmittance is $\eta_{\U{ch}}=10^{-0.2D/10}$ with $D$ 
denoting the fibre length. 
The overall misalignment error of the optical system is fixed to be $e_{\U{mis}}=1\%$. 
In addition, we assume an error correction leakage $\lambda_{\U{EC}}=f_{\U{EC}}|Z_{k_{\U{s}}}| h(e_z)$, 
where $e_z$ is the bit error rate of the sifted key $(Z_{{\rm A}}, Z_{{\rm B}})$. 
Moreover, for simplicity, we consider that the error correction efficiency of the protocol is a constant number 
$f_{\U{EC}}$=1.16 which does not depend on the size of $Z_{k_{\U{s}}}$. 
For simplicity, we model the imperfection of Alice's (Bob's) phase modulator as 
$\Delta\theta_{\U{A}}=\xi\theta_{\U{A}}/\pi$ ($\Delta\theta_{\U{B}}=-\Delta\theta_{\U{A}}$). 
Also, we consider that the intensity fluctuation of the laser source lies in the interval $[k^-, k^+]$ with 
$k^-=(1-r)k$ and $k^+=(1+r)k$ for a fixed value $r$. 

In these conditions, we simulate the secret key generation rate $R=\ell/N$ for a fixed value of the correctness coefficient 
$\epsilon_{\U{c}}=10^{-15}$. 
For this, we perform a numerical optimisation of the resulting secure key rate over the free parameters 
$p_z, p_{k_{\U{s}}},p_{k_{\U{d1}}}, k_{\U{s}}$ and $k_{\U{d1}}$. 

\subsection{A. Key generation rate for the exact intensity control case}
The resulting secret key rate for this scenario, {\it i.e.} when $r=0$, is shown in Fig. \ref{fig:keyrate1}. 
The security parameter is} $\epsilon_{\U{sec}}=10^{-10}$ and the total number of signals sent by Alice is 
$N=10^s$ with $s=9,10,11$ and $12$. 
We consider two possible cases: $\xi=0$~({\it i.e.}, the perfect encoding case) and $\xi=0.147$, which is equivalent to 
a phase modulation error of $8.42^{\circ}$. 
For comparison, Fig. \ref{fig:keyrate1} also includes the asymptotic secret key rate ({\it i.e.,} 
the key rate in the limit of infinitely large keys) with two decoy settings. 

As a result, we find that the effect of state preparation flaws on the key generation rate is almost negligible. 
Also, we have that if the total number of signals sent by Alice is about $N=10^{12}$, 
Alice and Bob can exchange secret keys over 150 km both when $\xi=0$ and $\xi=0,147$. 

Finally, Fig.~\ref{fig:block} shows the postprocessing block size $|Z_{k_{\U{s}}}|$ which is the length of the bit string to be 
processed in error correction and privacy amplification as a function of the distance when $N=10^s$ with $s=9,10,11$ and $12$. 

\subsection{B. Key generation rate for the intensity-fluctuation case}
In this section we evaluate the resulting secret key rate when the laser source suffers from intensity fluctuations. 
We study two cases: $r=0.02$ and $r=0.05$, 
where $r$ is the deviation rate from the expected value of the intensity. 
The results are shown in Figs.~\ref{fig:keyrate2} and~\ref{fig:keyrate3}. Here we consider that $N=\{10^{14}, 10^{15}\}$, 
and the term $\xi$ takes again the values $\xi=0$ and $\xi=0.147$. 
The security parameter is $\epsilon_{\U{sec}}=10^{-10}$ in Fig.~\ref{fig:keyrate2} and 
$\epsilon_{\U{sec}}=10^{-8}$ in Fig.~\ref{fig:keyrate3}. 

For comparison, these two figures also show the asymptotic secret key rate when Alice and Bob use two decoy settings. 
In this asymptotic case, we find that the degradation on the achievable key rate, 
when compared to the scenario $r=0$, is only about 10 km (20 km) when $r=0.02$ ($r=0.05$). 

In the finite-key regime, however, we obtain that the presence of intensity fluctuations seems to strongly limit 
the key generation rate if Alice and Bob do not know their probability distribution 
but only know the interval where the fluctuations lie in. 
For instance, when $N=10^{11}$ and $r=0$ (see Fig.~\ref{fig:keyrate1}) Alice and Bob can distribute a secret key 
over more than $100$ km. 
However, to achieve a similar secret key rate performance when the intensity fluctuation of the source is $2\%$ 
({\it i.e.}, the parameter $r=0.02$) they need to exchange about $N=10^{15}$ signals. 
The main technical reason for this behaviour seems to be the fact that Azuma's inequality~\cite{azuma} has a relatively 
slow convergence speed when compared to the Chernoff bound~\cite{chernoff} and the Multiplicative Chernoff bound~\cite{mdi:t2}.
 
As a side remark, let us mention that when $r=0.05$ and $N=10^{14}$ we find that the achievable secret key rate 
is basically zero unless we increase the security parameter $\epsilon_{\U{sec}}$ from $\epsilon_{\U{sec}}=10^{-10}$ 
to $\epsilon_{\U{sec}}=10^{-8}$. 
This is illustrated in Fig.~\ref{fig:keyrate3}.

Finally, Figs. \ref{fig:PPB1} and \ref{fig:PPB2} show the postprocessing block size $|Z_{k_{\U{s}}}|$ 
as a function of the distance when $N=10^s$ with $s=\{14,15\}$ for the 2\% intensity fluctuation case and for 
the 5\% intensity fluctuation case, respectively.

\section{V\hspace{-.1em}I. conclusion}
In summary, we have provided explicit security bounds for the loss-tolerant QKD protocol in the finite-key regime. 
On the application front, our results constitute an important step towards practical QKD with imperfect light sources, 
in that the resulting security performance is robust against encoding inaccuracies like, for instance, optical misalignments. 
Furthermore, our results take into account intensity fluctuations in the light source, 
which is a common experimental fact. 
Our results highlight the importance of the stable control of the intensity modulator as well as the need for a  
precise estimation of its intensity, which is not often sufficiently emphasised in the experiments. 
On a more general outlook, it would be of great practical interest to 
incorporate our results into measurement-device-independent QKD (mdiQKD)~\cite{mdi}. 
\section{V\hspace{-.1em}I\hspace{-.1em}I. acknowledgements}
We thank Lo H-K, Xu F, Kato G, Azuma K, Ikuta R, Nagamatsu Y,  Takeuchi Y and Matsuki K for valuable discussions on 
the security analysis and parameter estimation procedure, 
and Yoshino K, Fujiwara M, Sasaki M, and Tomita A for fruitful discussions on practical QKD systems. 
AM acknowledges support from the JSPS Grant-in-Aid for Scientific Research(A) 25247068. 
MC thanks the Galician Regional Government 
(program "Ayudas para proyectos de investigacion desarrollados por investigadores emergentes", 
and consolidation of Research Units: AtlantTIC), and the Spanish Government (project TEC2014-54898-R) for financial support. 
KT acknowledges support from the National Institute of Information and Communications Technology (NICT) of Japan 
(Project ``Secure Photonic Network Technology'' as part of Project UQCC) and the ImPACT program. 

\section{appendix a: derivation of the security bound}
\label{secA}
Here we present the calculations for the security bound given by equation (\ref{keyrate}). 
The security analysis is based on the universal composable security framework~\cite{composable,composable2}. 

Recall that after privacy amplification, the joint state shared by Alice, 
Bob and Eve is described by the following classical-quantum state
\begin{align} 
{\rho}^\U{actual}_{S_{\U{A}}S_{\U{B}}{\U{E}}}
=\sum_{s_{\U{A}},s_{\U{B}}}p(s_{\U{A}},s_{\U{B}})\ket{s_{\U{A}},s_{\U{B}}}\bra{s_{\U{A}},s_{\U{B}}}_{S_{\U{A}}S_{\U{B}}}
\otimes{\rho}^{s_{\U{A}},s_{\U{B}}}_{\U{E}},
\end{align}
where $s_{\U{A}}$ and $s_{\U{B}}$ are the classical bit strings for the keys, associated with orthonormal states $\ket{s_{\U{A}}}$ and 
$\ket{s_{\U{B}}}$ in a Hilbert space. Here, $p(s_{\U{A}}, s_{\U{B}})$ denotes 
the distribution of the keys and ${\rho}^{s_{\U{A}},s_{\U{B}}}_{\U{E}}$ is 
the quantum state of Eve's system conditioned on $S_{\U{A}}=s_{\U{A}}$ and $S_{\U{B}}=s_{\U{B}}$. In the ideal scenario, 
the joint state is described by 
\begin{align}
{\rho}^{\U{ideal}}_{S_{\U{A}}S_{\U{B}}\U{E}}=\frac{1}{2^{|s|}}\sum_{s}\ket{s,s}\bra{s,s}_{S_{\U{A}}S_{\U{B}}}\otimes{\rho}_{\U{E}},
\end{align}
where $S_{\U{A}}=S_{\U{B}}=s$ and ${\rho}_{\U{E}}$ is an arbitrary quantum state held by Eve. 
Using the security definition stated in the main text, a $\epsilon_\U{sec}$-secure QKD protocol satisfies 
\begin{align}
\frac{1}{2}\left\|{\rho}^\U{actual}_{S_{\U{A}}S_{\U{B}}\U{E}}-{\rho}^{\U{ideal}}_{S_{\U{A}}S_{\U{B}}\U{E}}\right \|_1\leq \epsilon_{\U{sec}}.
\label{composability}
\end{align} 
\newline Furthermore, if the security parameter $\epsilon_\U{sec}$ is appropriately chosen, it can be seen as the sum of errors in the correctness and secrecy, i.e., $\epsilon_\U{sec}=\epsilon_\U{s}+\epsilon_\U{c}$. To see this, let us introduce an intermediate state,
\begin{align}
{\rho}^{\U{inter}}_{S_{\U{A}}S_{\U{A}}{\U{E}}}=\sum_{s_{A}} p(s_{\U{A}})
\ket{s_{\U{A}},s_{\U{A}}}\bra{s_{\U{A}},s_{\U{A}}}_{S_{\U{A}}S_{\U{A}}}\otimes{\rho}^{s_{\U{A}}}_{\U{E}},
\end{align}
which is just a trivial classical extension of Alice's state. Then, by using the triangle inequality property of the trace distance metric, we have
\[
\frac{1}{2}\left\|{\rho}^\U{actual}_{S_{\U{A}}S_{\U{B}}{\U{E}}}-{\rho}^{\U{ideal}}_{S_{\U{A}}S_{\U{B}}\U{E}}\right 
\|_1\leq \frac{1}{2}\left\|{\rho}^\U{actual}_{S_{\U{A}}S_{\U{B}}{\U{E}}}-{\rho}^{\U{inter}}_{S_{\U{A}}S_{\U{A}}{\U{E}}} 
\right \|_1 + \frac{1}{2}\left\|{\rho}^{\U{inter}}_{S_{\U{A}}S_{\U{A}}{\U{E}}}-{\rho}^{\U{ideal}}_{S_{\U{A}}S_{\U{B}}\U{E}}\right \|_1.
\]
Fixing the first term on the R.H.S to $\epsilon_\U{c}$ gives
\[
\epsilon_\U{c}=\frac{1}{2}\left\|{\rho}^\U{actual}_{S_{\U{A}}S_{\U{B}}\U{E}}-{\rho}^{\U{inter}}_{S_{\U{A}}S_{\U{A}}{\U{E}}} \right \|_1 
\geq \frac{1}{2}\left\|{\rho}^\U{actual}_{S_{\U{A}}S_{\U{B}}}-{\rho}^{\U{inter}}_{S_{\U{A}}S_{\U{A}}} \right \|_1 = \Pr[S_{\U{A}} \not = S_{\U{B}}],
\]
where the inequality is due to the fact that the trace distance metric is contractive under any trace-preserving operation 
(in our case, the partial trace operation). Similarly, by fixing the second term to $\epsilon_\U{s}$ we have
\[
\epsilon_\U{s}=\frac{1}{2}\left\|{\rho}^{\U{inter}}_{S_{\U{A}}S_{\U{A}}{\U{E}}}-{\rho}^{\U{ideal}}_{S_{\U{A}}S_{\U{B}}\U{E}}\right \|_1 \geq 
\frac{1}{2}\left\|{\rho}^{\U{inter}}_{S_{\U{A}}{\U{E}}}-{\rho}^{\U{ideal}}_{S_{\U{A}}\U{E}}\right \|_1 .
\]
Therefore, fixing $\epsilon_\U{sec}=\epsilon_\U{s}+\epsilon_\U{c}$ gives the desired decomposition.

From \cite{koashi}, the lower bound on the secret key length of our protocol is written as
\begin{align}
\ell\geq\Big\lfloor{}m^{\U{L}}_0+m^{\U{L}}_1[1-\Gamma]-\lambda_{\U{EC}}-\log_2\frac{2}{\epsilon_{\U{c}}}\Big\rfloor,
\end{align}
where $m^{\U{L}}_0$ and $m^{\U{L}}_1$ are the lower bounds on the detection events of the vacuum and the single-photon emission, 
respectively, 
and $m^{\U{L}}_1\Gamma=m^{\U{L}}_1(h(e^{\U{U}}_{\U{ph}})+\delta)$ is the number of rounds performing the random hashing to 
correct the phase error, 
which is equivalent to the number of bits sacrificed in the privacy amplification step of the protocol. 
The parameter $\lambda_{\U{EC}}$ denotes the number of bits consumed in bit error correction, and 
$\lceil\log_2{1}/\epsilon_{\U{c}}\rceil\leq\log_2{2}/\epsilon_{\U{c}}$ is the 
length of the hash that Alice sends to Bob for the error verification using the universal$_2$ hash functions. 
From \cite{tsurumaru} and \cite{koashi07}, we have that $\epsilon_{\U{s}}$ can be bounded by 
$\epsilon_{\U{s}}\leq\sqrt{1-(1-\eta)(1-2^{-m^{\U{L}}_1\delta+1})}\leq{}
\sqrt{\eta+2^{-m^{\U{L}}_1\delta+1}}$. 
Therefore, the secret key length is obtained as
\begin{align}
\ell\geq\Big\lfloor{}m^{\U{L}}_0+m^{\U{L}}_1[1-h(e^{\U{U}}_{\U{ph}})]-\log_2\frac{2}{\epsilon^2_{\U{s}}-\eta}
-\lambda_{\U{EC}}-\log_2\frac{2}{\epsilon_{\U{c}}}\Big\rfloor,
\end{align}
where we consider $\eta$ as a fixed value in this paper. 

\section{appendix b: technical lemmas}
In this Appendix we introduce four different concentration inequalities which are used throughout this paper. 
First, we introduce the stochastic model that is assumed in {\it Lemmas 1, 2} and {\it 3}. 
\\

\noindent\textit{Stochastic model in {\it Lemmas 1, 2} and {\it 3}.}
\newline
Let $X_1, X_2...,X_N$ be a set of independent Bernoulli random variables 
that satisfy $\U{P}(X_i=1)=p_i$, and let $X:=\sum_{i=1}^NX_i$. 
The expected value of $X$ is denoted as $\mu:=E[X]=\sum_{i=1}^Np_i$. An observed outcome of $X$ is represented as $x$.
\newline
\newline
{\it Lemma~1.}~~\textit{Chernoff bound}~\cite{chernoff}
\newline
This bound requires the knowledge of $\mu$. It relates $x$ with $\mu$ as
\begin{align}
x=\mu+\delta_{\U{C}},
\label{cher}
\end{align}
except with error probability $\epsilon_{\U{C}}+\hat{\epsilon}_{\U{C}}$, 
where the fluctuation term $\delta_{\U{C}}$ lies in the interval $\delta_{\U{C}}\in{}[-\Delta_{\U{C}},\hat{\Delta}_{\U{C}}]$ with 
$\Delta_{\U{C}}=g_{\U{C}}(\mu,\epsilon_{\U{C}})$ and $\hat{\Delta}_{\U{C}}=\hat{g}_{\U{C}}(\mu,\hat{\epsilon}_{\U{C}})$, 
where $g_{\U{C}}(x,y)=\sqrt{2x\ln1/y}$ and $\hat{g}_{\U{C}}(x,y)=\sqrt{3x\ln1/y}$. 
Here the parameter $\epsilon_{\U{C}}~(\hat{\epsilon}_{\U{C}})$ denotes the probability that 
$x<\mu-\Delta_{\U{C}}~(x>\mu+\hat{\Delta}_{\U{C}})$. 
Equation~(\ref{cher}) holds if both 
$0<g_{\U{C}}(1/\mu,\epsilon_{\U{C}})<1$ and $0<\hat{g}_{\U{C}}(1/\mu,\hat{\epsilon}_{\U{C}})<1$ are met.
\\
\\{\it Lemma~2.}~~\textit{Hoeffding bound}~\cite{hoeff}\\
This bound does not require the knowledge of $\mu$. It relates $\mu$ and $x$ as
\begin{align}
\mu=x+\delta_{\U{H}},
\label{eq:hoeffding}
\end{align}
except with error probability $\epsilon_{\U{H}}+\hat{\epsilon}_{\U{H}}$, 
where the fluctuation term $\delta_{\U{H}}$ lies in the interval $\delta_{\U{H}}\in{}[-\Delta_{\U{H}},\hat{\Delta}_{\U{H}}]$ with 
$\Delta_{\U{H}}=g_{\U{H}}(N,\epsilon_{\U{H}})$ and $\hat{\Delta}_{\U{H}}=g_{\U{H}}(N,\hat{\epsilon}_{\U{H}})$, and where 
$g_{\U{H}}(x,y)=\sqrt{x/2\ln1/y}$.\\
\\{\it Lemma~3.}~~\textit{Multiplicative Chernoff bound}~\cite{mdi:t2}\\
This bound does not require the knowledge of $\mu$. It combines {\it Lemma~1} and {\it 2} above. It uses 
{\it Lemma 2} to estimate a lower bound on $\mu$ that is then basically used in combination with {\it Lemma 1}. 
In particular, let $\mu_{\rm L}=x-\sqrt{N/2\ln 1/\epsilon_{\U{H}}}$ for certain $\epsilon_{\U{H}}>0$. 
Then, if the following two conditions are satisfied: 
$(2\hat{\epsilon}_{\U{M}}^{-1})^{1/\mu_{{\rm L}}}\leq{\exp(9/32)}$ and $\epsilon_{\U{M}}^{-1/\mu_{{\rm L}}}<{\exp(1/3)}$ with 
$\hat{\epsilon}_{\U{M}},~\epsilon_{\U{M}}>0$, 
this Lemma states that $\mu$ and $x$ can be related as 
\begin{align}
\mu=x+\delta_{\U{M}},
\label{claim1}
\end{align}
except with error probability $\epsilon_{\U{H}}+\epsilon_{\U{M}}+\hat{\epsilon}_{\U{M}}$, 
where $\delta_{\U{M}}$ lies in the interval $[-{\Delta_M},\hat{\Delta}_M]$ with 
$\hat{\Delta}_{\U{M}}=g_{\U{M}}(x,\hat{\epsilon}_{\U{M}}^4/16)$ and $\Delta_{\U{M}}=g_{\U{M}}(x,{\epsilon_{\U{M}}}^{3/2})$, 
and where $g_{\U{M}}(x,y)=\sqrt{2x\ln(y^{-1})}$.
\newline
\newline 
{\it Lemma~4.~}\textit{Azuma's inequality}~\cite{azuma,azumaqkd}\\
This result plays a crucial role in our security analysis. 
It is applicable to any stochastic model as long as the Martingale and the Bounded difference conditions (BDC) are satisfied. 
In particular, a sequence of random variables $X^{(0)}, X^{(1)}, \ldots$ is called a Martingale if and only if 
$E[X^{(l+1)}|X^{(0)}, X^{(1)}, \ldots,X^{(l)}]=X^{(l)}$ for all non-negative integer $l$, 
where $E[\cdot]$ represents the expectation value. 
On the other hand, $X^{(0)}, X^{(1)}, \ldots$ is said to fulfil the BDC if there exists $c^{(l)}>0$ such that 
$|X^{(l+1)}-X^{(l)}|\leq{}c^{(l)}$ for all non-negative integer $l$. 

Let us consider $N$ trials of a random variable $X^{(l)}$, where $l$ refers to the $l$th trial. 
If $X^{(l)}$ is a Martingale and satisfies the BDC with $c^{(l)}=1$ then Azuma's inequality guarantees that
\begin{align}
\U{Pr}[|X^{(N)}-X^{(0)}|>N\delta]\leq{2e^{-\frac{N\delta^2}{2}}}
\label{eq:azuma1}
\end{align}
for any $\delta\in(0,1)$.

Let us  now define the following random variable for the $l$th trial,
\begin{align}
X^{(l)}:=\Lambda^{(l)}-\sum_{u=1}^lP(u|\xi_0, \ldots,\xi_{u-1}),
\label{eq:randomv}
\end{align}
where $\Lambda^{(l)}$ represents the actual number of events of the form $X^{(l)}=1$ observed among the first $l$ trials, 
and
$P(u|\xi_0,...,\xi_{u-1})$ is the conditional probability of having the event ``1'' in the $u$th trial conditioned on the 
first $u-1$ outcomes $\xi_0, \ldots,\xi_{u-1}$. In this scenario, it is straightforward to show that the random variables given by 
equation (\ref{eq:randomv}) are Martingale and satisfy the BDC with $c^{(l)}=1$. 
Hence, by applying Azuma's inequality we have that
\begin{align}
\U{Pr}[|\Lambda^{(N)}-\sum_{u=1}^NP(u|\xi_0,...,\xi_{u-1})|>N\delta]\leq{2e^{-\frac{N\delta^2}{2}}}.
\label{Azumainequality}
\end{align}
This means, in particular, that
\begin{align}
\Lambda^{(N)}=\sum_{u=1}^NP(u|\xi_0,...,\xi_{u-1})+\delta_{\U{A}}
\label{azumaequality}
\end{align}
except with error probability $\epsilon_\U{A}+\hat{\epsilon}_\U{A}$, where the parameter $\delta_\U{A}$ lies in the interval 
$\delta_{\U{A}}\in{}[-\Delta_{\U{A}},\hat{\Delta}_\U{A}]$ with $\Delta_\U{A}=g_\U{A}(N,\epsilon_\U{A})$ and 
$\hat{\Delta}_\U{A}=g_\U{A}(N,\hat{\epsilon}_\U{A})$, and where $g_\U{A}(x,y)=\sqrt{2x\ln(1/y)}$.

%AQUI

\section{appendix c: decoy-state analysis}
In this Appendix we first present the detail of the decoy-state analysis for the intensity fluctuation case and 
then we summarise all the equations for the decoy-state analysis, including those for the exact intensity control case. 
More precisely, we describe the estimation procedure that we use in order to obtain a lower bound on the number of 
vacuum contributions $T_{Z,0}$, and both a lower and an upper bound on the number of single-photon contributions 
$T_{Z,1}$. 

\subsection{1.~~Intensity fluctuation case}
Here, we generalise the decoy-state method to cover the case where the source suffers from intensity fluctuations. 
For this, as already mentioned previously, we shall consider that Alice and Bob only know the 
interval $[k^-, k^+]$ where the intensity $k$ lies.  

We begin by calculating the mean value $\langle{Z_k}\rangle$. 
Our starting point is the random variable $X^{(i|\overrightarrow{i-1})}_k$ for the $i$th trial 
when both Alice and Bob select the $Z$ basis. 
This random variable takes the value $1$ if Alice chooses the intensity $k$ and, moreover, 
the generated signal is detected by Bob; otherwise it is $0$. 
The term $\overrightarrow{i-1}$ 
reflects the fact that $X^{(i|\overrightarrow{i-1})}_k$ may depend on all the previous $i-1$ trials. 
With this notation, $\langle{Z_k}\rangle$ can be expressed as 
\begin{align}
\langle{Z_k}\rangle=\sum^{N_z}_{i=1}E[X^{(i|\overrightarrow{i-1})}_k]=\sum^{N_z}_{i=1}p^{(i|\overrightarrow{i-1})}(k\wedge\U{det}|Z),
\label{Zk}
\end{align}
where $N_z$ is the number of events where both Alice and Bob select the $Z$ basis. 
The probability $p^{(i|\overrightarrow{i-1})}(\ast)$ denotes the conditional probability that the event $\ast$ occurs in the $i$th trial 
conditioned on the results obtained in the previous $i-1$ trials, and 
the term $k\wedge\U{det}|Z$ represents the event where Alice selects the intensity $k$ 
and Bob detects the generated signal given that both of them have chosen the $Z$ basis. 
By using Bayes rule, we can rewrite equation (\ref{Zk}) as
\begin{align}
\langle{Z_k}\rangle&=p_k\sum^{N_z}_{i=1}\sum^\infty_{n=0}p^{(i|\overrightarrow{i-1})}(n\wedge\U{det}|k\wedge Z)\\
&=\frac{1}{p^2_{z}}\sum^{N_z}_{i=1}\sum^\infty_{n=0}p^{(i|\overrightarrow{i-1})}(k\wedge Z\wedge n\wedge\U{det})\\
&=\frac{1}{p^2_{z}}\sum^{N_z}_{i=1}\sum^\infty_{n=0}p^{(i)}(k\wedge Z\wedge n)p^{(i|\overrightarrow{i-1})}(\U{det}|k\wedge Z\wedge n)
\label{decoyproperty1}\\
&=p_k\sum^{N_z}_{i=1}\sum^\infty_{n=0}p^{(i)}(n|k\wedge Z)p^{(i|\overrightarrow{i-1})}(\U{det}|Z\wedge n), 
\label{decoyproperty2}
\end{align}
where $p_k$ is the probability that Alice chooses the intensity $k$; $n$ denotes an $n$-photon signal; 
$p_{z}$ represents the probability of selecting the $Z$ basis; and $p^{(i)}(\ast)$ is the probability that the event 
$\ast$ occurs in the $i$th trial. 
For instance, $p^{(i)}(n|k\wedge Z)$ is the conditional probability that Alice emits an $n$-photon state in the $i$th trial
given that she has chosen the intensity $k$ and both Alice and Bob have selected the $Z$ basis in the $i$th trial. 
Note that in the transformation from equation (\ref{decoyproperty1}) to equation (\ref{decoyproperty2}) 
we have used the property of the decoy-state method {\it i.e.,} 
$p^{(i|\overrightarrow{i-1})}(\U{det}|k\wedge Z\wedge n)=p^{(i|\overrightarrow{i-1})}(\U{det}|Z\wedge n)$.

In so doing, we obtain that $\langle{Z_k}\rangle$ is upper bounded by 
\begin{align}
\langle{Z_k}\rangle\leq p_k\sum^{N_z}_{i=1}\sum^\infty_{n=0}\frac{e^{-k^-}(k^+)^n}{n!} p^{(i|\overrightarrow{i-1})}(\U{det}|Z\wedge n).
\label{up}
\end{align}
\newline
Similarly, we find that 
\begin{align}
\langle{Z_k}\rangle&
\geq p_k\sum^{N_z}_{i=1}\sum^\infty_{n=0}\frac{e^{-k^+}(k^-)^n}{n!} p^{(i|\overrightarrow{i-1})}(\U{det}|Z\wedge n).
\label{low}
\end{align}
\newline
\textit{Lower bound on the number of vacuum contributions}
\newline
To obtain this bound, we first rewrite equations (\ref{up}) and (\ref{low}) for the cases 
$k=k_{\U{d2}}$ and $k=k_{\U{d1}}$, respectively. 
We obtain the following two inequalities,
\begin{align}
\frac{e^{k^-_{\U{d2}}}}{p_{k_{\U{d2}}}}\langle{Z_{k_{\U{d2}}}}\rangle\leq 
\sum^{N_z}_{i=1}p^{(i|\overrightarrow{i-1})}(\U{det}|0\wedge Z)+\sum^{N_z}_{i=1}p^{(i|\overrightarrow{i-1})}(\U{det}|1\wedge Z)k^+_{\U{d2}}
+\sum^{N_z}_{i=1}\sum_{n\geq 2}p^{(i|\overrightarrow{i-1})}(\U{det}|n\wedge Z)\frac{(k^+_{\U{d2}})^n}{n!}.
\label{kd2}
\end{align}
\begin{align}
     \frac{e^{k^+_{\U{d1}}}}{p_{k_{\U{d1}}}}\langle{Z_{k_{\U{d1}}}}\rangle\geq 
\sum^{N_z}_{i=1}p^{(i|\overrightarrow{i-1})}(\U{det}|0\wedge Z)+\sum^{N_z}_{i=1}p^{(i|\overrightarrow{i-1})}(\U{det}|1\wedge Z)k^-_{\U{d1}}
+\sum^{N_z}_{i=1}\sum_{n\geq 2}
p^{(i|\overrightarrow{i-1})}(\U{det}|n\wedge Z)\frac{(k^-_{\U{d1}})^n}{n!},
\label{kd1}
\end{align}
Next, we multiply equation (\ref{kd2}) by $k^-_{\U{d1}}$ and
equation (\ref{kd1}) by $k^+_{\U{d2}}$, and we add both 
expressions. 
In so doing, we find that
\begin{align}
(k^-_{\U{d1}}-k^+_{\U{d2}})\sum^{N_z}_{i=1}p^{(i|\overrightarrow{i-1})}(\U{det}|0\wedge Z)&\geq 
\frac{k^-_{\U{d1}}e^{k^-_{\U{d2}}}}{p_{k_{\U{d2}}}}\langle{Z_{k_{\U{d2}}}}\rangle-
\frac{k^+_{\U{d2}}e^{k^+_{\U{d1}}}}{p_{k_{\U{d1}}}}\langle{Z_{k_{\U{d1}}}}\rangle\nonumber\\
&+k^-_{\U{d1}}k^+_{\U{d2}}
\sum^{N_z}_{i=1}\sum_{n\geq2}\frac{(k^-_{\U{d1}})^{n-1}-(k^+_{\U{d2}})^{n-1}}{n!}p^{(i|\overrightarrow{i-1})}(\U{det}|n\wedge Z)\nonumber\\
&\geq k^-_{\U{d1}}\frac{e^{k^-_{\U{d2}}}}{p_{k_{\U{d2}}}}\langle{Z_{k_{\U{d2}}}}\rangle-k^+_{\U{d2}}
\frac{e^{k^+_{\U{d1}}}}{p_{k_{\U{d1}}}}\langle{Z_{k_{\U{d1}}}}\rangle,
\end{align}
where the second inequality holds because $k^-_{\U{d1}}>k^+_{\U{d2}}$. 

As a result, we find that $T_{Z,0}$ is lower bounded by
\begin{align}
T_{Z,0}:=\sum^{N_z}_{i=1}p^{(i|\overrightarrow{i-1})}(\U{det}|0\wedge Z)\geq \frac{1}{k^-_{\U{d1}}-k^+_{\U{d2}}}
\Big(k^-_{\U{d1}}\frac{e^{k^-_{\U{d2}}}}{p_{k_{\U{d2}}}}
\langle{Z_{k_{\U{d2}}}}\rangle-k^+_{\U{d2}}\frac{e^{k^+_{\U{d1}}}}{p_{k_{\U{d1}}}}\langle{Z_{k_{\U{d1}}}}\rangle\Big).
\label{TZ0}
\end{align}
To estimate the expectation values 
$\langle{Z_{k_{\U{d1}}}}\rangle$ and $\langle{Z_{k_{\U{d2}}}}\rangle$, 
we use Azuma's inequality because each trial of the random variables $X^{(i|\overrightarrow{i-1})}_{k_{\U{d1}}}$ and 
$X^{(i|\overrightarrow{i-1})}_{k_{\U{d2}}}$ may depend on the previous ones.
\newline
\newline
\textit{Lower bound on the number of single-photon contributions}
\newline
Here, we first particularise equations (\ref{up}) and (\ref{low}) for the cases $k=k_{\U{d1}}$ and $k=k_{\U{d2}}$, respectively. 
We have that
\begin{align}
     \frac{e^{k^-_{\U{d1}}}}{p_{k_{\U{d1}}}}\langle{Z_{k_{\U{d1}}}}\rangle
\leq \sum^{N_z}_{i=1}p^{(i|\overrightarrow{i-1})}(\U{det}|0\wedge Z)
+\sum^{N_z}_{i=1}p^{(i|\overrightarrow{i-1})}(\U{det}|1\wedge Z)(k^+_{\U{d1}})
+\sum_{n\geq 2}\sum^{N_z}_{i=1}p^{(i|\overrightarrow{i-1})}(\U{det}|n\wedge Z)\frac{(k^+_{\U{d1}})^n}{n!}.
\label{kd1sin}
\end{align}
\begin{align}
     \frac{e^{k^+_{\U{d2}}}}{p_{k_{\U{d2}}}}\langle{Z_{k_{\U{d2}}}}\rangle\geq 
\sum^{N_z}_{i=1}p^{(i|\overrightarrow{i-1})}(\U{det}|0\wedge Z)+\sum^{N_z}_{i=1}p^{(i|\overrightarrow{i-1})}(\U{det}|1\wedge Z)
(k^-_{\U{d2}})+\sum_{n\geq 2}\sum^{N_z}_{i=1}p^{(i|\overrightarrow{i-1})}(\U{det}|n\wedge Z)\frac{(k^-_{\U{d2}})^n}{n!},
\label{kd2sin}
\end{align}
Next, we add both expressions and we obtain
\begin{align}
&\frac{e^{k^+_{\U{d2}}}}{p_{k_{\U{d2}}}}\langle{Z_{k_{\U{d2}}}}\rangle
+\sum^{N_z}_{i=1}p^{(i|\overrightarrow{i-1})}(\U{det}|1\wedge Z)(k^+_{\U{d1}})
+\sum_{n\geq 2}\sum^{N_z}_{i=1}p^{(i|\overrightarrow{i-1})}(\U{det}|n\wedge Z)\frac{(k^+_{\U{d1}})^n}{n!}\nonumber\\
\geq&\frac{e^{k^-_{\U{d1}}}}{p_{k_{\U{d1}}}}\langle{Z_{k_{\U{d1}}}}\rangle
+\sum^{N_z}_{i=1}p^{(i|\overrightarrow{i-1})}(\U{det}|1\wedge Z)
(k^-_{\U{d2}})+\sum_{n\geq 2}\sum^{N_z}_{i=1}p^{(i|\overrightarrow{i-1})}(\U{det}|n\wedge Z)\frac{(k^-_{\U{d2}})^n}{n!}.
\end{align}
This last equation can be rewritten as 
\begin{align}
(k^+_{\U{d1}}-k^-_{\U{d2}})\sum^{N_z}_{i=1}p^{(i|\overrightarrow{i-1})}(\U{det}|1\wedge Z)\geq 
\frac{e^{k^-_{\U{d1}}}}{p_{k_{\U{d1}}}}\langle{Z_{k_{\U{d1}}}}\rangle-\frac{e^{k^+_{\U{d2}}}}{p_{k_{\U{d2}}}}
\langle{Z_{k_{\U{d2}}}}\rangle+\sum_{n\geq 2}\sum^{N_z}_{i=1}p^{(i|\overrightarrow{i-1})}(\U{det}|n\wedge Z)
\frac{(k^-_{\U{d2}})^n-(k^+_{\U{d1}})^n}{n!}.
\label{main}
\end{align}
Next, we evaluate the third term on the R.H.S of equation (\ref{main}). 
This term is lower bounded by
\begin{align}
\sum_{n\geq 2}\sum^{N_z}_{i=1}p^{(i|\overrightarrow{i-1})}(\U{det}|n\wedge Z)\frac{(k^-_{\U{d2}})^n-(k^+_{\U{d1}})^n}{n!}
&\geq \frac{(k^-_{\U{d2}})^2-(k^+_{\U{d1}})^2}{(k^-_{\U{s}})^2}
\sum_{n\geq 2}\sum^{N_z}_{i=1}p^{(i|\overrightarrow{i-1})}(\U{det}|n\wedge Z)\frac{(k^-_{\U{s}})^n}{n!},
\label{singleZ}
\end{align}
because when the conditions $n\geq2$, $k^+_{\U{d1}}>k^-_{\U{d2}}$ and $k^-_{\U{s}}>k^+_{\U{d1}}+k^-_{\U{d2}}$ are satisfied 
we have that 
$(k^-_{\U{d2}})^n-(k^+_{\U{d1}})^n\geq [(k^-_{\U{d2}})^2-(k^+_{\U{d1}})^2](k^-_{\U{s}})^{n-2}$. 
If we now use equation (\ref{low}) for $k=k_{\U{s}}$, we have that
\begin{align}
\sum_{n\geq 2}\sum^{N_z}_{i=1}p^{(i|\overrightarrow{i-1})}(\U{det}|n\wedge Z)\frac{(k^-_{\U{s}})^n}{n!}
\leq 
\frac{e^{k^+_{\U{s}}}}{p_{k_{\U{s}}}}\langle{Z_{k_{\U{s}}}}\rangle
-\sum^{N_z}_{i=1}\Big(p^{(i|\overrightarrow{i-1})}(\U{det}|0\wedge Z)+k^-_{\U{s}}p^{(i|\overrightarrow{i-1})}(\U{det}|1\wedge Z)\Big).
\label{cies}
\end{align}
The R.H.S of equation (\ref{singleZ}) can be lower bounded using the R.H.S of equation (\ref{cies}). 
This is so because $k^+_{\U{d1}} > k^-_{\U{d2}}$ and therefore the 
term $[(k^-_{\U{d2}})^2-(k^+_{\U{d1}})^2]/(k^-_{\U{s}})^2<0$ in equation (\ref{singleZ}). 
Hence, we have that the third term on the R.H.S of equation (\ref{main}) is lower bounded by 
\begin{align}
\sum_{n\geq 2}\sum^{N_z}_{i=1}p^{(i|\overrightarrow{i-1})}(\U{det}|n\wedge Z)
\frac{(k^-_{\U{d2}})^n-(k^+_{\U{d1}})^n}{n!}&\geq 
\frac{(k^-_{\U{d2}})^2-(k^+_{\U{d1}})^2}{(k^-_{\U{s}})^2}
\Big[
\frac{e^{k^+_{\U{s}}}}{p_{k_{\U{s}}}}\langle{Z_{k_{\U{s}}}}\rangle\nonumber\\
&-\sum^{N_z}_{i=1}\Big(p^{(i|\overrightarrow{i-1})}(\U{det}|0\wedge Z)+k^-_{\U{s}}p^{(i|\overrightarrow{i-1})}(\U{det}|1\wedge Z)\Big)
\Big].
\label{thirdlower}
\end{align}
That is, if we now combine equations (\ref{main}) and (\ref{thirdlower}) we find that
\begin{eqnarray}
(k^+_{\U{d1}}-k^-_{\U{d2}})\Big\{\frac{k^-_{\U{s}}-(k^+_{\U{d1}}+k^-_{\U{d2}})}{k^-_{\U{s}}}\Big\}
\sum^{N_z}_{i=1}p^{(i|\overrightarrow{i-1})}(\U{det}|1\wedge Z)&\geq& 
\frac{e^{k^-_{\U{d1}}}}{p_{k_{\U{d1}}}}\langle{Z_{k_{\U{d1}}}}\rangle-\frac{e^{k^+_{\U{d2}}}}{p_{k_{\U{d2}}}}\langle{Z_{k_{\U{d2}}}}\rangle\nonumber\\
&-&\frac{(k^+_{\U{d1}})^2-(k^-_{\U{d2}})^2}{(k^-_{\U{s}})^2}
\Big(\frac{e^{k^+_{\U{s}}}}{p_{k_{\U{s}}}}\langle{Z_{k_{\U{s}}}}\rangle-T^{\U{L}}_{Z,0}\Big),
\end{eqnarray}
which directly gives us a lower bound on $T_{Z,1}$,
\begin{align}
T_{Z,1}:&=\sum^{N_z}_{i=1}p^{(i|\overrightarrow{i-1})}(\U{det}|1\wedge Z)\nonumber\\
&\geq 
\frac{k^-_{\U{s}}}{(k^+_{\U{d1}}-k^-_{\U{d2}})(k^-_{\U{s}}-k^+_{\U{d1}}-k^-_{\U{d2}})}
\Big[\frac{e^{k^-_{\U{d1}}}}{p_{k_{\U{d1}}}}\langle{Z_{k_{\U{d1}}}}\rangle-\frac{e^{k^+_{\U{d2}}}}{p_{k_{\U{d2}}}}\langle{Z_{k_{\U{d2}}}}\rangle
-\frac{(k^+_{\U{d1}})^2-(k^-_{\U{d2}})^2}{(k^-_{\U{s}})^2}
\Big(\frac{e^{k^+_{\U{s}}}}{p_{k_{\U{s}}}}\langle{Z_{k_{\U{s}}}}\rangle-T^{\U{L}}_{Z,0}\Big)\Big].
\label{TZ1}
\end{align}

\noindent\textit{Upper bound on the number of single-photon contributions}
\newline
By adding equations (\ref{kd2}) and (\ref{kd1}), we have that 
\begin{align}
(k^-_{\U{d1}}-k^+_{\U{d2}})\sum^{N_z}_{i=1}p^{(i|\overrightarrow{i-1})}(\U{det}|1\wedge Z)&\leq 
\frac{e^{k^+_{\U{d1}}}}{p_{k_{\U{d1}}}}\langle{Z_{k_{\U{d1}}}}\rangle-
\frac{e^{k^-_{\U{d2}}}}{p_{k_{\U{d2}}}}\langle{Z_{k_{\U{d2}}}}\rangle
+\sum_{n\geq 2}\sum^{N_z}_{i=1}p^{(i|\overrightarrow{i-1})}(\U{det}|n\wedge Z)\frac{(k^+_{\U{d2}})^n-(k^-_{\U{d1}})^n}{n!}\nonumber\\
&\leq \frac{e^{k^+_{\U{d1}}}}{p_{k_{\U{d1}}}}\langle{Z_{k_{\U{d1}}}}\rangle-
\frac{e^{k^-_{\U{d2}}}}{p_{k_{\U{d2}}}}\langle{Z_{k_{\U{d2}}}}\rangle,
\label{eq:upS1}
\end{align}
where the second inequality holds because $k^-_{\U{d1}}>k^+_{\U{d2}}$. 
This means, in particular, that $T_{Z,1}$ is upper bounded by 
\begin{align}
T_{Z,1}&=\sum^{N_z}_{i=1}p^{(i|\overrightarrow{i-1})}(\U{det}|1\wedge Z)
\leq \frac{1}{k^-_{\U{d1}}-k^+_{\U{d2}}}\Big(\frac{e^{k^+_{\U{d1}}}}{p_{k_{\U{d1}}}}\langle{Z_{k_{\U{d1}}}}\rangle-
\frac{e^{k^-_{\U{d2}}}}{p_{k_{\U{d2}}}}\langle{Z_{k_{\U{d2}}}}\rangle\Big).
\end{align}

\subsection{2.~~Summary of the decoy-state analysis}
Here, we summarise all the equations needed in the decoy-state method, 
including those for the exact intensity control case. 
\newline
\newline
\textit{Lower bound on the number of vacuum contributions}\\
Let $\underline{\U{Decoy}_0}(ay,by')$ denote a lower bound on the number of events where Alice generates a vacuum state 
using the signal intensity and the basis setting $a\in\{Z, X\}$ to encode a bit value $y\in\{0,1\}$, 
and Bob observes the bit value $y'\in\{0,1\}$ when he measures the received signal using the basis $b\in\{Z,X\}$. 
%This quantity is given by~\cite{finite}
\begin{align}
\underline{\U{Decoy}_0}(ay,by')&=
\frac{p^-(k_{\U{s}}\wedge0)}{k^-_{\U{d1}}-k^+_{\U{d2}}}
\Big(\frac{k^-_{\U{d1}}e^{k^-_{\U{d2}}}}{p_{k_{\U{d2}}}}
\langle{a^yb^{y'}}_{k_{\U{d2}}}^-\rangle
-\frac{k^+_{\U{d2}}e^{k^+_{\U{d1}}}}{p_{k_{\U{d1}}}}\langle{a^{y}b^{y'}}_{k_{\U{d1}}}^+\rangle\Big), 
\end{align}
where the parameters $\langle{a^yb^{y'}}_{k_{\U{d2}}}^-\rangle$ and $\langle{a^{y}b^{y'}}_{k_{\U{d1}}}^+\rangle$ are defined 
in a similar way like equations (\ref{decoykd2vac}) and (\ref{decoykd1vac}) 
for the exact intensity control case and like equations (\ref{azumanumber1}) and (\ref{azumanumber2}) for 
the intensity-fluctuation case, respectively. 
The probability $p^-(k_{\U{s}}\wedge0)$ is a lower bound on $p(k_{\U{s}}\wedge 0)$ which denotes 
the probability that Alice selects the signal intensity setting and sends a vacuum state.
\newline
\newline
\textit{Lower bound on the number of single-photon contributions}
\newline
Let $\underline{\U{Decoy}_1}(ay,by')$ denote a lower bound on the number of events where Alice prepares a single-photon 
state using the signal intensity and the basis setting $a\in\{Z, X\}$ to encode a bit value $y\in\{0,1\}$, 
and Bob observes the bit value $y'\in\{0,1\}$ when he measures the received signal using the basis $b\in\{Z,X\}$. 
%This parameter has the form~\cite{finite}
\begin{align}
\underline{\U{Decoy}_1}(ay,by')&=
\frac{p^-(k_{\U{s}}\wedge1)k^-_{\U{s}}}{(k^+_{\U{d1}}-k^-_{\U{d2}})
(k^-_{\U{s}}-k^+_{\U{d1}}-k^-_{\U{d2}})}\Big[
\frac{e^{k^-_{\U{d1}}}}{p_{k_{\U{d1}}}}\langle {a^yb^{y'}}_{k_{\U{d1}}}^-\rangle
-\frac{e^{k^+_{\U{d2}}}}{p_{k_{\U{d2}}}}\langle{a^yb^{y'}}_{k_{\U{d2}}}^+\rangle\nonumber\\
&+\frac{(k^+_{\U{d1}})^2-(k^-_{\U{d2}})^2}{(k^-_{\U{s}})^2}
\Big(\frac{\underline{\U{Decoy}_0}(ay,by')}{p^-(k_{\U{s}}\wedge 0)}
-\frac{e^{k_{\U{s}}}\langle{a^yb^{y'}}_{k_{\U{s}}}^+\rangle}{p_{k_{\U{s}}}}\Big)
\Big],
\label{Eq:siglegain}
\end{align}
where the probability $p^-(k_{\U{s}}\wedge 1)$ is a lower bound on $p(k_{\U{s}}\wedge 1)$ 
which denotes the probability that Alice selects the signal intensity setting and sends a single-photon state.
\newline
\newline
\textit{Upper bound on the number of single-photon contributions}
\newline
Let $\overline{\U{Decoy}_1}(ay,by')$ denote an upper bound on the number of events where Alice prepares a single-photon state 
using the signal intensity and the basis setting $a\in\{Z, X\}$ to encode a bit value $y\in\{0,1\}$, 
and Bob observes the bit value $y'\in\{0,1\}$ when he measures the received signal using the basis $b\in\{Z,X\}$. 
%This parameter is given by~\cite{finite}
\begin{align}
\overline{\U{Decoy}_1}(ay,by')&=\frac{p^+(k_{\U{s}}\wedge1)}{k^-_{\U{d1}}-k^+_{\U{d2}}}
\Big(\frac{e^{k^+_{\U{d1}}}}{p_{k_{\U{d1}}}}\langle{a^yb^{y'}}_{k_{\U{d1}}}^+\rangle
-\frac{e^{k^-_{\U{d2}}}}{p_{k_{\U{d2}}}}\langle{a^yb^{y'}}_{k_{\U{d2}}}^-\rangle\Big),
\label{uppersingle}
\end{align}
where the probability $p^+(k_{\U{s}}\wedge1)$ is an upper bound on $p(k_{\U{s}}\wedge1)$.

\section{Appendix D: PHASE ERROR RATE ESTIMATION}
In this Appendix we explain how to derive equation (\ref{Nphase}). That is, we obtain 
an upper bound on the number of phase errors associated to the single-photon pulses emitted by Alice when she 
selects the signal intensity setting, both Alice and Bob use the $Z$ basis, 
and Bob obtains a successful detection event ({\it i.e.}, $y'\neq\emptyset$). 
As we are interested in the phase error rate defined in the single-photon emission events and all the statistics associated 
with the single-photons can be estimated using the decoy state method, in the virtual protocol we only consider the cases 
where Alice emits single photons.

For this, in the security proof we consider a virtual protocol that based on the complementarity argument~\cite{s8} 
is equivalent to the actual protocol. 
In the virtual scheme, Alice prepares an ancilla qubit which is entangled with the pulse that she sends to Bob. 
Importantly, from Eve's viewpoint both protocols are completely indistinguishable because they emit the 
same quantum states and announce the same classical information. 

Let us start our analysis by introducing the following joint states, 
which we shall denote as $\ket{\tilde{\psi}_{j_z}}_{\U{A_1,B}}$. 
They are a purification of the signals $\tilde{{\rho}}_{jz}$ with $j\in\{0,1\}$ (see equation (\ref{tilderho})), 
\begin{align}
\ket{\tilde{\psi}_{j_z}}_{\U{A_1,B}}=\sqrt{P^{j_z}_0}\ket{0}_{\U{A_1}}\ket{\phi^{j_z}_{0}}_{\U{B}}
+\sqrt{P^{j_z}_1}\ket{1}_{\U{A_1}}\ket{\phi^{j_z}_{1}}_{\U{B}},
\label{purestate}
\end{align}
where the index $\U{A_1}$ represents the ancilla system and the index B is the system that Alice sends to Bob. 
In addition, we define the state:
\begin{align}
\ket{\tilde{\Psi}_{z}}_{\U{A_1,A_2,B}}=\frac{1}{\sqrt{2}}
\Big(\ket{0}_{\U{A_2}}\ket{\tilde{\psi}_{0_z}}_{\U{A_1,B}}+\ket{1}_{\U{A_2}}\ket{\tilde{\psi}_{1_z}}_{\U{A_1,B}}\Big),
\label{virstate1}
\end{align}
where the ancilla system $\U{A_2}$ stores the bit information. 
The aim of the virtual protocol is to quantify how accurately Bob can estimate Alice's measurement outcome 
if she would measure system $\U{A_2}$ in the complementarity basis 
({\it i.e.}, if she would use the POVM ${M}_{X,\U{A_2}}=\{\ket{+}\bra{+}, \ket{-}\bra{-}\}$, 
where $\ket{\pm}=1/\sqrt{2}(\ket{0}\pm\ket{1})$). 
This way one can characterise the information that Eve could have obtained about the raw key~\cite{s8}. 
Note that equation (\ref{virstate1}) can be rewritten as
\begin{align}
\ket{\tilde{\Psi}_{z}}_{\U{A_1,A_2,B}}=
\sqrt{\frac{1+\langle{\tilde{\psi}_{0_z}}|\tilde{\psi}_{1_z}\rangle_{\U{A_1,B}}}{2}}\ket{+}_{\U{A_2}}
\ket{\tilde{\psi}^{\U{vir}}_{0_x}}_{\U{A_1,B}}
+\sqrt{\frac{1-\langle{\tilde{\psi}_{0_z}}|\tilde{\psi}_{1_z}\rangle_{\U{A_1,B}}}{2}}\ket{-}_{\U{A_2}}
\ket{\tilde{\psi}^{\U{vir}}_{1_x}}_{\U{A_1,B}},
\label{virstate2}
\end{align}
where the normalised virtual states $\ket{\tilde{\psi}^{\U{vir}}_{j_x}}_{\U{A_1,B}}$, with $j\in\{0,1\}$, are defined as 
\begin{align}
\ket{\tilde{\psi}^{\U{vir}}_{j_x}}_{\U{A_1,B}}=\frac{\ket{\tilde{\psi}_{0_z}}_{\U{A_1,B}}
+(-1)^j\ket{\tilde{\psi}_{1_z}}_{\U{A_1,B}}}
{\sqrt{2\big[1+(-1)^j\langle{\tilde{\psi}_{0_z}}|\tilde{\psi}_{1_z}\rangle_{\U{A_1,B}}}\big]}.
\label{ksem}
\end{align}

Let us now introduce some additional notation before we describe in detail the different steps of the virtual protocol. 
In particular, the states prepared by Alice in the virtual protocol are given by
\begin{align}
\ket{\phi}_{\U{sh,A_1,B}}=\sum_{c=1}^5\sqrt{P(c)}\ket{c}_{\U{sh}}\ket{\phi^{(c)}}_{\U{A_1,B}},
\label{mv}
\end{align}
where the shield system sh belongs to Alice's laboratory, the states $\ket{\phi^{(c)}}_{\U{A_1,B}}$ have the form
\begin{align}
\ket{\phi^{(1)}}_{\U{A_1,B}}=\ket{\tilde{\psi}^{\U{vir}}_{0_x}}_{\U{A_1,B}},\nonumber\\ 
\ket{\phi^{(2)}}_{\U{A_1,B}}=\ket{\tilde{\psi}^{\U{vir}}_{1_x}}_{\U{A_1,B}}, \nonumber\\
\ket{\phi^{(3)}}_{\U{A_1,B}}=\ket{\tilde{\psi}_{0_z}}_{\U{A_1,B}}, \nonumber\\
\ket{\phi^{(4)}}_{\U{A_1,B}}=\ket{\tilde{\psi}_{1_z}}_{\U{A_1,B}}, \nonumber\\
\ket{\phi^{(5)}}_{\U{A_1,B}}=\ket{\tilde{\psi}_{0_x}}_{\U{A_1,B}},
\label{phiOmega}
\end{align}
and the probabilities $P(c)$ are given by
\begin{align}
P(1)&=\frac{p_z^2}{2}\left(1+\langle{\tilde{\psi}_{0_z}}|{\tilde{\psi}_{1_z}}\rangle_{\U{A_1,B}}\right),\nonumber\\
P(2)&=\frac{p_z^2}{2}\left(1-\langle{\tilde{\psi}_{0_z}}|{\tilde{\psi}_{1_z}}\rangle_{\U{A_1,B}}\right),\nonumber\\
P(3)&=P(4)=\frac{p_zp_x}{2},\nonumber\\
P(5)&=p_x.
\label{Prob}
\end{align}
Also, we define Bob's POVM for the $Z$ and the $X$ basis measurement as ${M}_{Z,\U{B}}=\{{M}_{Z0}, {M}_{Z1}, {M}_{Z\U{f}}\}$ and 
${M}_{X,\U{B}}=\{{M}_{X\U{0}}, {M}_{X\U{1}}, {M}_{X\U{f}}\}$, respectively. 
Here, the operator ${M}_{Z(X)\U{f}}$ corresponds to the inconclusive outcome in the $Z~(X)$ basis. 
Importantly, in the security analysis we assume that this operator is the same for both basis, 
{\it i.e.,} ${M}_{\U{f}}:={M}_{Z\U{f}}={M}_{X\U{f}}$. 
This assumption allows us to conceptually delay Bob's measurement basis choice until he is certain to obtain a 
conclusive result. 
That is, we can consider that Bob first conducts a filter operation with Kraus operators 
$D=\{\sqrt{I-{M}_{\U{f}}}, \sqrt{{M}_{\U{f}}}\}$ 
followed by the $Z$ or $X$ basis measurement, 
which we redefine as $\{M_{Z0}, M_{Z1}\}$ and $\{M_{X0}, M_{X1}\}$, respectively. 

Next we present the steps of the virtual protocol in detail. 
\newline
\begin{breakbox}
{\it {\textbf{Virtual protocol}}}
\newline
Alice repeats the first step $n_1$ times, where $n_1$ is the number of single-photon emissions generated by Alice 
in the actual protocol within the set $|Z_{k_{\U{s}}}|$. 
\newline
1.~~{\it Preparation}
\newline
Alice prepares the state $\ket{\phi}_{\U{sh,A_1,B}}$ given by equation~(\ref{mv}). 
Afterwards, she sends Bob system B over a quantum channel and delays her measurement on system sh until step 3. 
\newline
\newline
2.~~{\it Filter operation}
\newline
Bob performs on system B the filter operation $D$ and, if this operation succeeds, he stores this system in a quantum memory. 
We will denote the set of successful filter results as $\mbox{\boldmath $\U{S}$}$, and $|$$\mbox{\boldmath $\U{S}$}$$|=N_1$. 
\newline
\newline
3.~~{\it Collective measurement}
\newline
Alice and Bob perform on the states in the set $\mbox{\boldmath $\U{S}$}$ a collective measurement characterised by the POVM 
elements ${F}_{\Omega,s}$, with $\Omega\in\{1,2,...,6\}$ and $s\in\{0,1\}$ (see equation~(\ref{cm1})) 
on the states in the set $\mbox{\boldmath $\U{S}$}$.
\newline
\newline
4.~~{\it Classical communication}
\newline
Alice announces the $Z~(X)$ basis choice over an authenticated public channel when the result of her measurement in step 3 
is $\Omega=1,2,3,4~(\Omega=5,6)$. 
Then, Bob announces the $Z~(X)$ basis choice, also over an authenticated public channel, when the measurement outcome in step 3 
is $\Omega=1,2,6~(\Omega=3,4,5)$ to ensure that the classical information declared in both the actual 
and the virtual protocols coincide (see the main text below for further details). 
In addition, Bob declares the value of $s$ when $\Omega=3,4,5,6$.
\newline
\newline
5.~~{\it Estimation of the number of phase errors}
\newline
Alice and Bob calculate an upper bound on the number of phase errors. This upper bound is given by
\begin{align}
N^{\U{U}}_{\U{ph}}=\overline{\Lambda^{(N_1)}_{1,1}}+\overline{\Lambda^{(N_1)}_{{2,0}}},
\label{eph}
\end{align}
where $\Lambda^{(N_1)}_{\Omega,{s}}$ denotes the number of outcomes associated to the operator ${F}_{\Omega,s}$ after 
$N_1$ trials, and $\overline{\Lambda^{(N_1)}_{\Omega,{s}}}$ 
is an upper bound on $\Lambda^{(N_1)}_{\Omega,{s}}$. 
\end{breakbox}
\quad

The size of the set $\mbox{\boldmath $\U{S}$}$ (see step 2 of the virtual protocol) is upper bounded by
\begin{align}
|\mbox{\boldmath $\U{S}$}|=N_1\leq{}\sum_{a,b\in\{Z,X\}}\sum_{y,y'\in\{0,1\}}\overline{\U{Decoy_1}}(ay,by'),
\end{align} 
where the parameter $\overline{\U{Decoy_1}}(ay,by')$ is defined in Appendix C. 
Also, the POVM elements ${F}_{\Omega,s}$ of Alice and Bob's collective measurement are given by
\begin{align}
{F}_{\Omega,s}&={P}[\ket{\Omega}_{\U{sh}}]\otimes{}{M}_{X\U{s}} \quad \quad \textrm{when $\Omega\in\{1,2,3,4\}$,}\nonumber \\
{F}_{5,s}&={P}[\ket{5}_{\U{sh}}]\otimes{}p_x{M}_{Xs},\nonumber \\
{F}_{6,s}&={P}[\ket{5}_{\U{sh}}]\otimes{}p_z{M}_{Zs}.
\label{cm1}
\end{align}
These POVM elements satisfy $\sum_{s\in\{0,1\}}\sum_{\Omega\in\{1,...,6\}}{F}_{\Omega,s}={I}_{\U{sh}}\otimes{}I_{\U{B}}$. 

It is easy to demonstrate that from Eve's viewpoint the virtual protocol described above is completely equivalent 
to the actual protocol. 
Indeed, the quantum states that Alice sends to Bob are exactly the same in both protocols. 
Also, both schemes declare precisely the same classical information. 
To see this last point, let us further clarify the fourth step of the virtual protocol. 
In particular, note that when $\Omega=1 (2)$ the state that Alice sends to Bob in the virtual protocol is 
$\U{Tr}_{\U{A_1}}P[\ket{\tilde{\psi}_{0(1)x}}_{\U{A_1B}}]$ and Bob uses the $X$ basis. 
However, in this case, Alice and Bob announce the $Z$ basis. 
In so doing, the actual and virtual protocols are indistinguishable. 
This is so because in the actual protocol the events $\Omega=1$ or $2$ are used to generate a secret key, 
${\it i.e.,}$ in these events both Alice and Bob select, and therefore also declare, the $Z$ basis. 
Then, the virtual protocol has to do the same declaration, otherwise it could be distinguished from the actual protocol. 
That is, with our definition of the virtual protocol we guarantee that it produces precisely 
the same classical information as the actual protocol. 

Next, we present the estimation method that we use in order to upper bound the quantities 
$\Lambda^{(N_1)}_{{1,1}}$ and $\Lambda^{(N_1)}_{{2,0}}$ using experimentally observed values. 
For this, we consider the sequence of random variables $X^{(l)}_{{\Omega, {s}}}$, with $l=1,...,N_1$, given by 
\begin{align}
X^{(l)}_{{\Omega, {s}}}=\Lambda^{(l)}_{{\Omega,{s}}}-\sum_{u=1}^lP_{{\Omega,{s}}}(u|\xi_0,...,\xi_{u-1}),
\label{eq:randomvariable}
\end{align}
where $P_{{\Omega,{s}}}(u|\xi_0,...,\xi_{u-1})$ is the conditional probability of obtaining the values 
$\Omega$ and $s$ in the collective measurement performed in the $u$th trial of the third step of the virtual protocol, 
conditioned on the first $u-1$ measurement outcomes from the collective measurements $\xi_0,...,\xi_{u-1}$. 
To obtain this conditional probability we use the following joint state in $N_1$ trials, 
\begin{align}
\ket{\Phi}_{\U{sh,A_1,B}}=\ket{\phi_{\overrightarrow{u-1}}}_{\U{sh,A_1,B}}\ket{\phi_u}_{\U{sh,A_1,B}}
\ket{\phi_{\overrightarrow{N_1-u}}}_{\U{sh,A_1,B}},
\end{align}
where $\ket{\phi_{\overrightarrow{u-1}}}_{\U{sh,A_1,B}}$, $\ket{\phi_u}_{\U{sh,A_1,B}}$, and 
$\ket{\phi_{\overrightarrow{N_1-u}}}_{\U{sh,A_1,B}}$ represent, respectively, Alice's prepared states 
in the first $u-1$ trials, in the $u$th trial, and in the rest of trials. 

Let $U_{\U{BE}}$ denote Eve's unitary transformation on Bob's system B and on her system E. We have that 
\begin{align}
{U}_{\U{BE}}\ket{\Phi}_{\U{sh,A_1,B}}\ket{0}_{\U{E}}=\sum_tB_{t,\U{B}}\ket{\Phi}_{\U{sh,A_1,B}}\ket{t}_{\U{E}},
\end{align}
where $B_{t,\U{B}}$ denotes the Kraus operator which acts on system B depending on Eve's measurement outcome of 
her ancilla. 
Now we consider Alice and Bob's collective measurement. 
In particular, let ${M}_{\U{sh}_v,s_v}$ represent the Kraus operator associated with the $v$th $(1\leq v\leq u)$ measurement outcome 
of Alice's system sh and Bob's system. 
Also, let $\mathcal{O}_{{u}-1,\U{sh,B}}$ denote Alice and Bob's joint measurement operator up to $u-1$ trials. 
It can be written as 
\begin{align}
\mathcal{O}_{u-1,\U{sh,B}}=\bigotimes_{v=1}^{u-1}{M}_{\U{sh}_v,s_v}(I_{\U{sh}}\otimes{}\sqrt{1-{M}_{\U{f}}}).
\end{align}
We shall denote the measurement outcomes of the first $u-1$ trials as $O_{u-1}$. 
Then, after Eve's intervention and conditioned on the fact of obtaining the measurement results $O_{u-1}$, 
we have that the normalised $u$th state of Alice's system sh and Bob's system B, which we shall represent as 
${\rho}^{\U{sh,B}}_{u|\overrightarrow{u-1}}$, is given by 
\begin{align}
{\rho}^{\U{sh,B}}_{u|\overrightarrow{u-1}}
&=\frac{{\sigma}^{\U{sh,B}}_{u|O_{u-1}}}{\U{Tr}\left({\sigma}^{\U{sh,B}}_{u|O_{u-1}}\right)},
\end{align}
where the state ${\sigma}^{\U{sh,B}}_{u|O_{u-1}}$ has the form 
\begin{align}
{\sigma}^{\U{sh,B}}_{u|O_{u-1}}&:=\sum_t\U{Tr}_{\bar{u}}\Big({P}[\mathcal{O}_{u-1,\U{sh,B}}{B}_{t,\U{B}}\ket{\Phi}_{\U{sh,A_1,B}}]\Big).
\label{eq:sigma}
\end{align}
Here, $\U{Tr}_{\bar{u}}$ is the trace over all systems except for the $u$th systems sh and B. 
Equation~(\ref{eq:sigma}) can be rewritten as follows:
\begin{align}
{\sigma}^{\U{sh,B}}_{u|O_{u-1}}&=
\sum_t\sum_{\overrightarrow{u-1},\overrightarrow{N_1-u}}\U{Tr}^{(u)}_{\U{A_1}}\Big(P[\bra{\overrightarrow{u-1}}
\bra{\overrightarrow{N_1-u}}\mathcal{O}_{u-1,\U{sh,B}}
B_{t,\U{B}}\ket{\phi_{\overrightarrow{u-1}}}_{\U{sh,A_1,B}}{\ket{\phi_u}_{\U{sh,A_1,B}}}\ket{\phi_{\overrightarrow{N_1-u}}}_{\U{sh,A_1,B}}
]\Big)\nonumber\\
&=\sum_t\sum_{\overrightarrow{u-1},\overrightarrow{N_1-u}}\U{Tr}^{(u)}_{\U{A_1}}\Big(P[
A^{\overrightarrow{u-1},\overrightarrow{N_1-u}}_{t,\U{B}|O_{u-1}}\ket{\phi_u}_{\U{sh,A_1,B}}]\Big),
\end{align}
where $\U{Tr}^{(u)}_{\U{A_1}}$ represents the trace over the $u$th $\U{A_1}$ system, the states 
$\ket{\overrightarrow{u-1}}$ and $\ket{\overrightarrow{N_1-u}}$ denote an orthogonal 
basis for the first $u-1$ systems and the last $N_1-u$ systems, respectively, and 
\begin{align}
{A}^{\overrightarrow{u-1},\overrightarrow{N_1-u}}_{t,\U{B}|O_{u-1}}:=\bra{\overrightarrow{u-1}}
\bra{\overrightarrow{N_1-u}}\mathcal{O}_{u-1,{\U{sh,B}}}{B}_{t,\U{B}}
\ket{\phi_{\overrightarrow{u-1}}}_{\U{sh,A_1,B}}\ket{\phi_{\overrightarrow{N_1-u}}}_{\U{sh,A_1,B}}
\label{eq:A}
\end{align}
is the Kraus operator acting on the $u$th system conditioned on the measurement outcomes $O_{u-1}$. 

Therefore, we obtain that the conditional probability defined in equation (\ref{eq:randomvariable}) for 
$\Omega\in\{1,...,6\}$ is given by 
\begin{align}
P_{{\Omega,{s}}}(u|\xi_0,...,\xi_{u-1})&=\U{Tr}\left\{{F}_{\Omega,s}\frac{\mathcal{E}({\rho}^{\U{sh,B}}_{u|\overrightarrow{u-1}})}
{\U{Tr}\left[\mathcal{E}(\rho^{\U{sh,B}}_{u|\overrightarrow{u-1}})\right]}\right\}\nonumber\\
&=\frac{Q(\Omega)}{\U{Tr}\left[\mathcal{E}({\rho}^{\U{sh,B}}_{u|\overrightarrow{u-1}})\right]\U{Tr}({\sigma}^{\U{sh,B}}_{u|O_{u-1}})}\U{Tr}
\left[\mathcal{M}^{(u|\overrightarrow{u-1})}_{Xs}\U{Tr}_{\U{A_1}}P[\ket{\phi^{(\Omega)}}_{\U{A_1,B}}]\right]\nonumber\\
&=:Q(\Omega)\U{T}^{(u|\overrightarrow{u-1})}_{M_{Xs}}\Big[\U{Tr}_{\U{A_1}}{P}[\ket{\phi^{(\Omega)}}_{\U{A_1,B}}]\Big],
\label{def:P}
\end{align}
where $\mathcal{E}({\rho}):=({I_{\U{sh}}}\otimes\sqrt{{I}-{M}_{\U{f}}}){\rho}({I_{\U{sh}}}\otimes\sqrt{{I}-{M}_{\U{f}}})^{\dagger}$, 
the probability $Q(\Omega)=P(\Omega)$ for $\Omega\in\{1,2,3,4\}$, $Q(5)=p_xP(5)$ and $Q(6)=p_zP(5)$,
the operator 
$\mathcal{M}^{(u|\overrightarrow{u-1})}_{Xs}:=\sum_t\sum_{\overrightarrow{u-1},\overrightarrow{N_1-u}}
(\sqrt{{I}-{M}_{\U{f}}}~{A}^{\overrightarrow{u-1},\overrightarrow{N_1-u}}_{t,\U{B}|O_{u-1}})^{\dagger}
M_{Xs}(\sqrt{{I}-{M}_{\U{f}}}~{A}^{\overrightarrow{u-1},\overrightarrow{N_1-u}}_{t,\U{B}|O_{u-1}})$, 
the states $\ket{\phi^{(\Omega)}}_{\U{A_1,B}}$ are defined in equation~(\ref{phiOmega}), 
and $\U{T}^{(u|\overrightarrow{u-1})}_{M_{Xs}}\Big[\U{Tr_{A_1}}P[\ket{\phi^{(\Omega)}}_{\U{A_1,B}}]\Big]$ is the 
$u$th conditional probability that Bob's measurement outcome in the $X$ basis is $s\in\{0,1\}$ given that 
Alice sends him the state $\U{Tr_{A_1}}P[\ket{\phi^{(\Omega)}}_{\U{A_1,B}}]$ and the filter operation succeeds conditioned on 
the first $u-1$ measurement results. 
For convenience, we shall refer to $\U{T}^{(u|\overrightarrow{u-1})}_{M_{Xs}}[A]$ as the transmission rate of $A$.

If we now apply Azuma's inequality (see {\it Lemma 4} in Appendix B), we obtain 
\begin{align}
&\Big|\sum_{u=1}^{N_1}P_{{\Omega,s}}(u|\xi_{0},...,\xi_{u-1})-\Lambda^{(N_1)}_{{\Omega,s}}\Big|\leq\Delta^s_{\U{A},\Omega},
\label{azumaineq}
\end{align}
except with error probability $\epsilon^s_{\U{A},\Omega}$, where 
$\Delta^s_{\U{A},\Omega}=g_{\U{A}}(N_1,\epsilon^s_{\U{A},\Omega})$. 

By combining this result with that from equation (\ref{def:P}), we have that
\begin{align}
\frac{\Lambda^{(N_1)}_{{\Omega,s}}-\Delta^s_{\U{A},\Omega}}{Q(\Omega)}&\leq{}
\sum_{u=1}^{N_1}\U{T}^{(u|\overrightarrow{u-1})}_{M_{Xs}}
\Big[\U{Tr}_{\U{A_1}}P[\ket{\phi^{(\Omega)}}_{\U{A_1,B}}]\Big]
\leq\frac{\Lambda^{(N_1)}_{{\Omega,s}}+\Delta^s_{\U{A},\Omega}}{Q(\Omega)}.
\label{three}
\end{align}
Note that the parameters $\Lambda^{(N_1)}_{{\Omega,s}}$, with $\Omega\in\{3,4,5\}$, can be 
upper and lower bounded using the decoy-state method. 
We shall denote the failure probability of this estimation as $\epsilon_{Z0,Xs}$, $\epsilon_{Z1,Xs}$ and $\epsilon_{X0,Xs}$, 
respectively. 
 
As a result, we obtain bounds on 
$\sum_{u=1}^{N_1}\U{T}^{(u|\overrightarrow{u-1})}_{M_{Xs}}\Big[\U{Tr_{A_1}}P[\ket{\phi^{(\Omega)}}_{\U{A_1,B}}]\Big]$
that maximise the number of phase errors $N_{\U{ph}}$ in the single-photon 
emissions within the set $|Z_{k_{\U{s}}}|$. 
They are denoted as $N_{M_{Xs}}(\Omega)$ and have the form
\begin{align}
N_{M_{Xs}}(3):=\Big\{\frac{\underline{\U{Decoy}_1}(Z0,Xs)-\Delta^s_{\U{A},3}}{Q(3)}~\U{or}~
\frac{\overline{\U{Decoy}_1}(Z0,Xs)+\Delta^s_{\U{A},3}}{Q(3)}\Big\},
\label{azuma3}
\end{align}
\begin{align}
N_{M_{Xs}}(4):=\Big\{\frac{\underline{\U{Decoy}_1}(Z1,Xs)-\Delta^s_{\U{A},4}}{Q(4)}~\U{or}~
\frac{\overline{\U{Decoy}_1}(Z1,Xs)+\Delta^s_{\U{A},4}}{Q(4)}\Big\},
\label{azuma4}
\end{align}
\begin{align}
N_{M_{Xs}}(5):=\Big\{\frac{\underline{\U{Decoy}_1}(X0,Xs)-\Delta^s_{\U{A},5}}{Q(5)}~\U{or}~
\frac{\overline{\U{Decoy}_1}(X0,Xs)+\Delta^s_{\U{A},5}}{Q(5)}\Big\}.
\label{azuma5}
\end{align}

When $\Omega\in\{1,2\}$, the quantity 
$\U{T}^{(u|\overrightarrow{u-1})}_{M_{Xs\oplus{1}}}\Big[\tilde{\rho}^{\U{vir}}_{sx}\Big]$, with $s\in\{0,1\}$, represents 
the transmission rate of the virtual states 
$\tilde{\rho}^{\U{vir}}_{sx}=\U{Tr_{B}}(P[\ket{\tilde{\psi}^{\U{vir}}_{s_x}}_{\U{A_1,B}}])$, 
with $\ket{\tilde{\psi}^{\U{vir}}_{s_x}}_{\U{A_1,B}}$ given by equation~(\ref{ksem}). 
This quantity can be decomposed into the transmission rate of the Pauli operators 
$\sigma_{I}, \sigma_{X}$ and $\sigma_{Z}$. 
However, for later convenience, we will decompose it as a function of $\tilde{\rho}_{0z}$ and $\tilde{\rho}_{1z}$, 
together with $\sigma_{I}, \sigma_{X}$ and $\sigma_Z$. 
Here, the states $\tilde{\rho}_{0z}$ and $\tilde{\rho}_{1z}$ are defined in equation (\ref{filteredstate}). 
In particular, from equation (\ref{ksem}) we find that 
\begin{align}
\U{T}^{(u|\overrightarrow{u-1})}_{M_{Xs\oplus{1}}}\Big[\tilde{\rho}^{\U{vir}}_{sx}\Big]
&=\frac{1}{2\big[1+(-1)^s\langle{\tilde{\psi}_{0_z}}|\tilde{\psi}_{1_z}\rangle_{\U{A_1,B}}\big]}
\Biggl\{
\U{T}^{(u|\overrightarrow{u-1})}_{M_{Xs\oplus{1}}}[\tilde{\rho}_{0z}]+\U{T}^{(u|\overrightarrow{u-1})}_{M_{Xs\oplus{1}}}[\tilde{\rho}_{1z}]
+(-1)^s\Big(\U{T}^{(u|\overrightarrow{u-1})}_{M_{Xs\oplus{1}}}\Big[\U{Tr_B}(P[\ket{\tilde{\psi}_{0_z}}\bra{\tilde{\psi}_{1_z}}_{\U{A_1,B}}])\Big]
\nonumber\\
&
+\U{T}^{(u|\overrightarrow{u-1})}_{M_{Xs\oplus{1}}}\Big[\U{Tr_B}(P[\ket{\tilde{\psi}_{1_z}}\bra{\tilde{\psi}_{0_z}}_{\U{A_1,B}}])\Big]\Big)\Biggr\}
\nonumber\\
&=\frac{1}{2\{1+(-1)^s(\sqrt{P^{0z}_{0}P^{1z}_{0}}\expect{\phi^{0z}_0|\phi^{1z}_0}+
\sqrt{P^{0z}_{1}P^{1z}_{1}}\expect{\phi^{0z}_1|\phi^{1z}_1})\}}\Big[
\U{T}^{(u|\overrightarrow{u-1})}_{M_{Xs\oplus{1}}}[\tilde{\rho}_{0z}]
+\U{T}^{(u|\overrightarrow{u-1})}_{M_{Xs\oplus{1}}}[\tilde{\rho}_{1z}]\nonumber\\
&+(-1)^s\sum^1_{t=0}\sqrt{P^{0z}_{t}P^{1z}_{t}}\Big\{(a^{0z}_ta^{1z}_t+b^{0z}_tb^{1z}_t)
\U{T}^{(u|\overrightarrow{u-1})}_{M_{Xs\oplus{1}}}[\sigma_I]+
(a^{0z}_tb^{1z}_t+a^{0z}_tb^{1z}_t)\U{T}^{(u|\overrightarrow{u-1})}_{M_{Xs\oplus{1}}}[\sigma_{X}]\nonumber\\
&+(a^{0z}_ta^{1z}_t-b^{0z}_tb^{1z}_t)\U{T}^{(u|\overrightarrow{u-1})}_{M_{Xs\oplus{1}}}[\sigma_{Z}]
\Big],
\label{virtualdecomposition}
\end{align}
where we have used equation (\ref{purestate}) in the second equality and see equation (\ref{eigenvec}) for the 
definition of $a^S_t$ and $b^S_t$.

In addition, we have that the transmission rate of $\tilde{\rho}_{0z}, \tilde{\rho}_{1z}$ and $\tilde{\rho}_{0x}$ can be decomposed 
using the Pauli operators as follows
\begin{align}
&\left( \begin{array}{ccc} \U{T}^{(u|\overrightarrow{u-1})}_{M_{Xs{\oplus{1}}}}[\tilde{\rho}_{0z}]
\\ \U{T}^{(u|\overrightarrow{u-1})}_{M_{Xs{\oplus{1}}}}[\tilde{\rho}_{1z}]\\ 
\U{T}^{(u|\overrightarrow{u-1})}_{M_{Xs{\oplus{1}}}}
[\tilde{\rho}_{0x}]
\\ \end{array} \right) 
=\left(\begin{array}{ccc} 1/2 & r^{0z}_{x}/2 & r^{0z}_{z}/2 
\\ 1/2 & r^{1z}_{x}/2 &r^{1z}_{z}/2\\ 1/2 & r^{0x}_{x}/2 &
r^{0x}_{z}/2\\\end{array} \right) 
\left( \begin{array}{ccc} \U{T}^{(u|\overrightarrow{u-1})}_{M_{Xs{\oplus{1}}}}[\sigma_I]\\ 
\U{T}^{(u|\overrightarrow{u-1})}_{M_{Xs{\oplus{1}}}}
[\sigma_X]\\\U{T}^{(u|\overrightarrow{u-1})}_{M_{Xs{\oplus{1}}}}[\sigma_Z]\\ 
\end{array} \right)=:A\left( \begin{array}{ccc} \U{T}^{(u|\overrightarrow{u-1})}_{M_{Xs{\oplus{1}}}}[\sigma_I]\\ 
\U{T}^{(u|\overrightarrow{u-1})}_{M_{Xs{\oplus{1}}}}[\sigma_{X}]\\\U{T}^{(u|\overrightarrow{u-1})}_{M_{Xs{\oplus{1}}}}[\sigma_{Z}]\\ 
\end{array} \right).
\end{align}
Hence, the transmission rate of the Pauli operators can be described as 
\begin{align}
\left( \begin{array}{ccc} \U{T}^{(u|\overrightarrow{u-1})}_{M_{Xs{\oplus{1}}}}[\sigma_I]\\ 
\U{T}^{(u|\overrightarrow{u-1})}_{M_{Xs{\oplus{1}}}}[\sigma_X]\\\U{T}^{(u|\overrightarrow{u-1})}_{M_{Xs{\oplus{1}}}}[\sigma_Z]\\ 
\end{array} \right)
=A^{-1}\left( \begin{array}{ccc} \U{T}^{(u|\overrightarrow{u-1})}_{M_{Xs{\oplus{1}}}}[\tilde{\rho}_{0z}]
\\ \U{T}^{(u|\overrightarrow{u-1})}_{M_{Xs{\oplus{1}}}}[\tilde{\rho}_{1z}]\\ 
\U{T}^{(u|\overrightarrow{u-1})}_{M_{Xs{\oplus{1}}}}
[\tilde{\rho}_{0x}]
\\ \end{array} \right),
\label{matrixrelation}
\end{align}
where the inverse matrix $A^{-1}$ is given in equation (\ref{matrixA}). 

Now, if we combine equations (\ref{virtualdecomposition}), (\ref{matrixrelation}) and (\ref{three}), we 
obtain that $N_{\U{ph}}$ is upper bounded by
\begin{align}
N_{\U{ph}}&=\Lambda^{(N_1)}_{1,1}+\Lambda^{(N_1)}_{2,0}\leq{}
\sum^1_{s=0}P(s+{1})\sum_{u=1}^{N_1}\U{T}^{(u|\overrightarrow{u-1})}_{M_{Xs\oplus{1}}}\Big[
\tilde{\rho}^{\U{vir}}_{sx}\Big]+\Delta^{s\oplus{1}}_{\U{A},s+1}\nonumber\\
&=\sum^1_{s=0}
\frac{P(s+{1})}{2\{1+(-1)^s(\sqrt{P^{0z}_{0}P^{1z}_{0}}\expect{\phi^{0z}_0|\phi^{1z}_0}+
\sqrt{P^{0z}_{1}P^{1z}_{1}}\expect{\phi^{0z}_1|\phi^{1z}_1})\}}\Bigg[
\sum^{N_1}_{u=1}\U{T}^{(u|\overrightarrow{u-1})}_{M_{Xs\oplus{1}}}[\tilde{\rho}_{0z}]
+\sum^{N_1}_{u=1}\U{T}^{(u|\overrightarrow{u-1})}_{M_{Xs\oplus{1}}}[\tilde{\rho}_{1z}]\nonumber\\
&+(-1)^s\sum^1_{t=0}\sqrt{P^{0z}_{t}P^{1z}_{t}}\Big\{C_{t,0}
\sum^{N_1}_{u=1}\U{T}^{(u|\overrightarrow{u-1})}_{M_{Xs\oplus{1}}}[\tilde{\rho}_{0z}]+
C_{t,1}\sum^{N_1}_{u=1}\U{T}^{(u|\overrightarrow{u-1})}_{M_{Xs\oplus{1}}}[\tilde{\rho}_{1z}]
+C_{t,2}\sum^{N_1}_{u=1}\U{T}^{(u|\overrightarrow{u-1})}_{M_{Xs\oplus{1}}}[\tilde{\rho}_{0x}]\Big\}\Bigg]+\Delta^{s\oplus 1}_{\U{A},s+1}.
\label{N_ph}
\end{align}

Finally, by using the results given by equations (\ref{azuma3})-(\ref{azuma5}), we find that
\begin{align}
N_{\U{ph}}&\leq{}\sum^1_{s=0}
\frac{P(s+1)}{2\{1+(-1)^s(\sqrt{P^{0z}_{0}P^{1z}_{0}}\expect{\phi^{0z}_0|\phi^{1z}_0}+
\sqrt{P^{0z}_{1}P^{1z}_{1}}\expect{\phi^{0z}_1|\phi^{1z}_1})\}}\Big[
N_{M_{Xs}}(3)+N_{M_{Xs}}(4)\nonumber\\
&+(-1)^s\sum^1_{t=0}\sqrt{P^{0z}_{t}P^{1z}_{t}}\Big\{C_{t,0}N_{M_{Xs}}(3)+C_{t,1}N_{M_{Xs}}(4)
+C_{t,2}N_{M_{Xs}}(5)\Big\}\Big]+\Delta^{s\oplus 1}_{\U{A},s+1}\\
&=:N^{\U{U}}_{\U{ph}},
\end{align}
except with error probability 
\begin{align}
\varepsilon_{\U{ph}}=\epsilon^1_{\U{A},1}+\epsilon^0_{\U{A},2}+\sum_{s\in\{0,1\},\Omega\in\{3,4,5\}}\epsilon^s_{\U{A},\Omega}
+\sum_{s\in\{0,1\}}(\epsilon_{Z0,Xs}+\epsilon_{Z1,Xs}+\epsilon_{X0,Xs}),
\label{epsilon}
\end{align}
%%Marcsos's commentf
where $\epsilon^s_{\U{A},\Omega}$ is the failure probability that equation (\ref{azumaineq}) does not hold for 
$\Omega\in\{1,...,5\}$ and $s\in\{0,1\}$. Also, $\epsilon_{Z0(1),Xs}$ and $\epsilon_{X0,Xs}$ are the failure probabilities
of the decoy state method {\it i.e.,} the failure probabilities of the estimation of $\Lambda^{(N_1)}_{3(4),s}$ and 
$\Lambda^{(N_1)}_{5,s}$, respectively. 
\section{Appendix e: simulation}
In this Appendix we present the calculations used to obtain Figs.~\ref{fig:keyrate1}, \ref{fig:keyrate2} and 
\ref{fig:keyrate3} in the main text. 

In particular, we consider that Alice sends Bob pairs of coherent states of the form 
$\ket{\sqrt{k_{\U{ref}}} e^{i\chi}}_{\U{r}}\ket{\sqrt{k_{\U{sig}}} e^{i(\chi+\theta_{\U{A}}+\Delta\theta_{\U{A}})}}_{\U{s}}$, 
and we set Alice's (Bob's) phase modulation error to $\Delta\theta_{\U{A}}=\xi\theta_{\U{A}}/\pi$ 
($\Delta_{\U{B}}=-\Delta_{\U{A}}$). 
Also, we assume a Gaussian distribution for the intensity fluctuations of the laser within an interval $[k^-,k^+]$. 
That is, we consider that the probability density function of the fluctuations 
is given by $p_{\U{G}}(k)=A\exp[-(k-\mu)^2/2\sigma^2]$, 
where $\mu$ is the desired value ({\it e.g.}, $k_{\U{s}}, k_{\U{d1}}$, and $k_{\U{d2}}$), the dispersion $\sigma^2$ has the form 
$\sigma^2=r\mu/5$, and the normalisation factor $A$ is such that $\int^{k^+}_{k^-}p_{\U{G}}(k)\U{d}k=1$.
 \newline
\newline
\textit{Calculation of the parameters $m_{0}^{{\rm L}}$ and $m_{1}^{{\rm L}}$}
\newline
For this, we need to obtain $|Z_{k}|$ for all $k\in{K}$. 
Afterwards, we simply apply the procedure described in section~IV.A (for the exact intensity control case) and 
in section~IV.B (for the intensity fluctuation case). 

We consider that the total number of pulses sent by Alice using the intensity setting $k$ is given by 
$N_{k}=Np_{k}$, where $N$ denotes the total number of transmissions until the conditions in the Sifting step of the protocol are met. 
The total system loss $\eta_{\U{sy}}:=\eta_{\U{det}}\eta_{\U{ch}}$ includes the channel loss and the detection efficiency 
of Bob's detectors. 
The conditional probability $p^{(k)}(Zj|Zi)$ 
that Bob obtains the bit $j\in\{0,1\}$ using the $Z$ basis given that 
Alice sends him a bit $i$ encoded with the intensity $k$ and also in the $Z$ basis can be written as 
\begin{align}
&p^{(k)}(Z0|Z0)=\int^{k^+}_{k^-} p_{\U{G}}(k)\Big[1-(1-p_{\U{d}})e^{-\eta_{\U{sy}}k}\Big]\U{d}k,\\
&p^{(k)}(Z1|Z0)=p_{\U{d}},\\
&p^{(k)}(Zj|Z1)=\int^{k^+}_{k^-} p_{\U{G}}(k)\Big[1-(1-p_{\U{d}})
\exp\Big(-\frac{\eta_{\U{sy}}k(1-(-1)^j\cos{\xi})}{2}\Big)\Big]\U{d}k.
\end{align} 
The conditional probability $p^{(k)}(Zj\wedge{}\overline{Zj\oplus{1}}|Zi)$ that Bob 
interprets the bit value $j$ (after a random assignment of double click events to single clicks events) when he uses the $Z$ basis 
given that Alice sends him a pulse with the intensity $k$, prepared in the $Z$ basis, 
and encoding the bit value $i$ is written as
\begin{align}
&p^{(k)}(Zj\wedge{\overline{Zj\oplus1}}|Zi)=p^{(k)}(Zj|Zi)(1-p^{(k)}(Zj\oplus1|Zi))
+\frac{1}{2}p^{(k)}(Zj|Zi)p^{(k)}(Zj\oplus1|Zi).
\label{sat2}
\end{align} 
To simulate the misalignment in the optical system we transform this probability as
\begin{align}
&P^{(k)}(Zj\wedge{\overline{Zj\oplus{}1}}|Zj)=p^{(k)}(Zj\wedge{\overline{Zj\oplus{}1}}|Zj)(1-e_{\U{mis}}),\\
&P^{(k)}(Zj\oplus{}1\wedge{\overline{Zj}}|Zj)=p^{(k)}(Zj\wedge{\overline{Zj\oplus{}1}}|Zj)e_{\U{mis}}
+p^{(k)}(Zj\oplus{}1\wedge{\overline{Zj}}|Zj). 
\end{align} 
In so doing, we obtain 
\begin{align}
|Z_{k}|=N_{k}p_z^2\sum_{i,j\in\{0,1\}}P^{(k)}(Zj\wedge{\overline{Zj\oplus{}1}}|Zi).
\end{align}
The bit error rate in the $Z$ basis when Alice sends Bob a pulse using the signal intensity is given by
\begin{align}
e_{z}=\frac{\sum_{j\in\{0,1\}}P^{(k_{\U{s}})}(Zj\oplus{1}\wedge{\overline{Zj}}|Zj)}{\sum_{i,j\in\{0,1\}}
P^{(k_{\U{s}})}(Zj\wedge{\overline{Zj\oplus{}1}}|Zi)}.
\end{align}
\textit{Calculation of the parameter $N_{\U{ph}}$}
\newline
According to equation (\ref{Nphase}), we have that $N_{\U{ph}}$ is upper bounded by
\begin{align}
N_{\U{ph}}\leq&\frac{1-\sin\frac{\xi}{2}}{2}\Big(\frac{p_z}{p_x}\Big)^2
(\overline{\U{Decoy}_1}(X0,X1)+\Delta^1_{\U{A},5})+
\frac{p_z}{p_x}(\overline{\U{Decoy}_1}(Z0,X0)+\overline{\U{Decoy}_1}(Z1,X0)+\Delta^0_{\U{A},3}+\Delta^0_{\U{A},4})\nonumber\\
&-\frac{1-\sin\frac{\xi}{2}}{2}\Big(\frac{p_z}{p_x}\Big)^2(\underline{\U{Decoy}_1}(X0,X0)+\Delta^0_{\U{A},5})
+\Delta_{\U{A},1}+\Delta_{\U{A},2}.
\end{align}
To obtain $\overline{\U{Decoy}_1}(X0,X1)$ and $\underline{\U{Decoy}_1}(X0,X0)$ we first calculate the probability 
$p^{(k)}(Xj\wedge{}\overline{Xj\oplus{1}}|X0)$ that Bob obtains the bit $j$ with the $X$ basis 
given that Alice sends him a pulse of intensity $k$ using the $X$ basis and encoding the bit value $0$. 
For this, we have that
\begin{align}
p^{(k)}(Xj|X0)=\int^{k^+}_{k^-} p_{\U{G}}(k)\Big[1-\exp[-\frac{\eta_{\U{sy}}k(1+(-1)^j\cos\xi)}{2}](1-p_{\U{d}})\Big]\U{d}k.
\end{align}
Then, by using equation (\ref{sat2}) we find
\begin{align}
p^{(k)}(Xj\wedge{\overline{Xj\oplus{}1}}|X0)=p^{(k)}(Xj|X0)(1-p^{(k)}(Xj\oplus1|X0))
+\frac{1}{2}p^{(k)}(X0|X0)p^{(k)}(X1|X0).
\end{align}
Finally, we include the effect of the misalignment in the optical systems. 
That is, we transform $p^{(k)}(Xj\wedge{}\overline{Xj\oplus{1}}|Xi)$ as
\begin{align}
&P^{(k)}(X0\wedge{\overline{X1}}|X0)=p^{(k)}(X0\wedge{\overline{X1}}|X0)(1-e_{\U{mis}}),\\
&P^{(k)}(X1\wedge{\overline{X0}}|X0)=p^{(k)}(X0\wedge{\overline{X1}}|X0)e_{\U{mis}}
+p^{(k)}(X1\wedge{\overline{X0}}|X0).
\end{align}
The number $|X^j_k|$ is therefore given by
\begin{align}
|X^j_k|=N_{k}p_x^2P^{(k)}(Xj\wedge{\overline{Xj\oplus{}1}}|X0).
\end{align}

Next, we calculate $\overline{\U{Decoy}_1}(Z0,X0)$ and 
$\overline{\U{Decoy}_1}(Z1,X0)$. 
For this we need to obtain $|Z^iX^j_k|$. 
We have that the probability $p^{(k)}(Xj|Zi)$ that Bob obtains the bit $j$ with the $X$ basis given that Alice sends him a pulse 
of intensity $k$, prepared in the $Z$ basis, and encoding the bit value $i$ is given by 
\begin{align}
&p^{(k)}(Xj|Z0)=\int^{k^+}_{k^-} 
p_{\U{G}}(k)\Big[1-\exp\Big(\frac{-\eta_{\U{sy}}k(1+(-1)^j\sin\xi/2)}{2}\Big)(1-p_{\U{d}})\Big]\U{d}k,\\
&p^{(k)}(Xj|Z1)=\int^{k^+}_{k^-} p_{\U{G}}(k)\Big[1-\exp\Big(\frac{-\eta_{\U{sy}}k(1-(-1)^j
\sin3\xi/2)}{2}\Big)(1-p_{\U{d}})\Big]\U{d}k.
\end{align}
In this scenario the probability 
$P^{(k)}(Xj\wedge{}\overline{Xj\oplus{1}}|Zi)$ has the form 
\begin{align}
P^{(k)}(Xj\wedge{}\overline{Xj\oplus{1}}|Zi)
=p^{(k)}(Xj|Zi)(1-p^{(k)}(Xj\oplus{1}|Zi))
+\frac{1}{2}p^{(k)}(Xj|Zi)p^{(k)}(Xj\oplus{1}|Zi),
\end{align}
and therefore the quantity $|Z^iX^j_k|$ can be written as
\begin{align}
|Z^iX^j_k|=N_{k}\frac{p_xp_z}{2}P^{(k)}(Xj\wedge{}\overline{Xj\oplus{1}}|Zi).
\end{align}

\end{document}